\documentclass[letterpaper, 10 pt, conference]{ieeeconf}
\IEEEoverridecommandlockouts
\usepackage{cite}
\usepackage{amsmath,amssymb,amsfonts}
\usepackage{algorithmic}
\usepackage{graphicx}
\usepackage{textcomp}
\usepackage{xcolor}
\usepackage{mathtools}
\usepackage{ntheorem}
\usepackage{epstopdf}
\usepackage{caption}
\usepackage{subcaption}
\usepackage{setspace}
\usepackage{subcaption}
\usepackage{dsfont}
\usepackage{cleveref}
\makeatletter
\renewcommand*\env@matrix[1][*\c@MaxMatrixCols c]{%
  \hskip -\arraycolsep
  \let\@ifnextchar\new@ifnextchar
  \array{#1}}
\makeatother

\newcommand{\Nat}{{\mathds N}} %Conjunto de los Naturales
{}
\newtheorem{corollary}{Corollary}{}
{}
{}
\newtheorem{theorem}{Theorem}{}
{}
\newtheorem{lemma}{Lemma}{}
{}

\def\BibTeX{{\rm B\kern-.05em{\sc i\kern-.025em b}\kern-.08em
    T\kern-.1667em\lower.7ex\hbox{E}\kern-.125emX}}
\begin{document}

\title{\LARGE \bf
Infinite Horizon Privacy in Networked Control Systems:\\ Utility/Privacy Tradeoffs and Design Tools 
}

\author{Haleh Hayati, Nathan van de Wouw, Carlos Murguia% <-this % stops a space
% <-this % stops a space
%\thanks{This work was partially supported by the Australian Research Council (ARC) under the Discovery Project DP170104099.}
\thanks{Haleh Hayati, Nathan van de Wouw, and Carlos Murguia are with the Department of Mechanical Engineering, Dynamics and Control Group, Eindhoven University of Technology, The Netherlands. Emails: \& h.hayati@tue.nl, \& n.v.d.wouw@tue.nl, \& c.g.murguia@tue.nl.}
}
\maketitle
\begin{abstract}
We address the problem of synthesizing distorting mechanisms that maximize infinite horizon privacy for Networked Control Systems (NCSs). We consider stochastic LTI systems where information about the system state is obtained through noisy sensor measurements and transmitted to a (possibly adversarial) remote station via unsecured/public communication networks to compute control actions (a remote LQR controller). Because the network/station is untrustworthy, adversaries might access sensor and control data and estimate the system state. To mitigate this risk, we pass sensor and control data through distorting (privacy-preserving) mechanisms before transmission and send the distorted data through the communication network. These mechanisms consist of a linear coordinate transformation and additive-dependent Gaussian vectors. We formulate the synthesis of the distorting mechanisms as a convex program. In this convex program, we minimize the infinite horizon mutual information (our privacy metric) between the system state and its optimal estimate at the remote station for a desired upper bound on the control performance degradation (LQR cost) induced by the distortion mechanism.
\end{abstract}
%%%%%%%%%%%%%%%%%%%%%%%%%%%%%%%%%%%%%%%%%%%%%%%%%%%%%%%%%%%%%%%%%%%
\section{Introduction}
In recent years, control systems have become increasingly distributed and networked. Networked Control Systems (NCSs) involve closing control loops over real-time communication networks. This allows controllers, sensors, and actuators to be connected through multipurpose networks, providing benefits such as increased system flexibility, ease of installation and maintenance, and decreased wiring and cost \cite{hespanha2007survey}. However, when estimation/control tasks in NCSs are performed by third parties %, and/or the communication network is public/unsecured, 
information sharing might result in private information leakage \cite{Takashi_1}-\nocite{Takashi_2}\nocite{murguia2020secure}\cite{Cao_Privacy}.
%these advantages come at the cost of security threats and privacy loss
%%%% figure for the system without privacy??

In NCSs, information about the plant state, say $x$, is obtained through sensor measurements and then sent through communication networks to a remote station to perform computations, e.g., estimation or control tasks. Shared information is correlated with private variables that carry sensitive information, e.g., the state itself (as it can reveal private system trajectories like reactant levels and user behavior, or it could be used to launch state-dependent attacks \cite{hayati2022privacy}), and references (because they can reveal manufactured products specs, tracked trajectories, and visited locations). If communication networks and/or the remote station are untrustworthy, adversaries might access and estimate the system state. To avoid this, we randomize the disclosed data before transmission using additive-dependent Gaussian random vectors and transmit the distorted data over the network. %, the measurement data that is transmitted from the plant to the remote station, and the controller that is transmitted from the remote station to the plant,

Using additive random noise is common practice to enforce privacy of sensitive data. In the context of privacy of databases, a popular approach is differential privacy \cite{Jerome1}, where random noise is added to the response of queries so that private information stored in the database cannot be inferred. Differential privacy has also been applied to various estimation and control problems \cite{Jerome1,cortes2016differential}. % However, most results using differential privacy do not consider the trade-off between privacy and utility, i.e., they do not account for the negative effect of the privacy mechanism on the application performance. In general, when considering privacy-utility trade-offs, optimal additive noise distributions are determined by the particular privacy and distortion metrics, and the system configuration \cite{Topcu}-\nocite{SORIA}\nocite{Geng}\cite{Dullerud}.
There are also techniques addressing privacy in dynamical systems from an information-theoretic perspective, see \cite{Poor1,Farokhi1,FAROKHI3,hayati2021finite}. In this line of work, privacy is characterized using information-theoretic metrics, e.g., mutual information, entropy, and Kullback-Leibler divergence. However, independently of the metric being used, if the data to be kept private follows continuous probability distributions, the problem of finding the optimal additive noise to maximize privacy is difficult to solve \cite{Farokhi1}. This issue has been addressed by assuming the data to be kept private is deterministic \cite{Farokhi1}. However, in a Cyber-Physical-Systems context, the inherent system dynamics and unavoidable system and sensor noise lead to stochastic non-stationary data, and thus, existing tools do not fit this problem setting. 

It is crucial to note that data privacy fundamentally differs between static data, like databases, and dynamically correlated data, e.g., in feedback control systems. In networked control architectures, information flows bidirectionally between the remote station and the plant. The authors in \cite{Takashi_2} demonstrate the necessity of privacy masks for information flow directions by identifying the infinite horizon privacy consequences of bidirectional information flow in feedback control. To the best of the authors' knowledge, no privacy-preserving design tools are offered for MIMO multidimensional feedback control systems that minimize infinite horizon bidirectional information flow while maintaining a desired closed-loop control performance. There are works addressing information-theoretic infinite-horizon privacy \cite{fang2021fundamental,Takashi_2} for SISO scalar systems. Also, in \cite{tanaka2017lqg,yazdani2022differentially}, the infinite horizon privacy is considered for MIMO multidimensional feedback control systems, but considering the information flow in one direction. 

Motivated by these results, in this manuscript, we present an optimization-based framework for synthesizing privacy-preserving Gaussian mechanisms that maximize privacy but keep distortion on control performance bounded. The proposed privacy mechanism consists of a coordinate transformation and additive Gaussian vectors that are designed to hide (as much as possible) the private state of the plant \cite{hayati2021finite}. We distort disclosed data in both information flow directions, the measurement data in the uplink direction that is transmitted from the plant to the remote station and the control data in the downlink direction that is transmitted from the remote station to the plant. We show that using coordinate transformations in the privacy mechanism (in combination with additive Gaussian vectors) can effectively reduce information leakage significantly more than adding stochastic vectors only. Note that it is not desired to overly distort the control performance while minimizing the information leakage. Therefore, when designing the privacy mechanisms, we consider the trade-off between \emph{privacy} and \emph{performance degradation}. As \emph{performance metric}, we use the \emph{LQR control cost} of the closed-loop system when operating on distorted privacy-preserving data. We follow an information-theoretic approach to privacy. As \emph{privacy metric}, we use the \emph{mutual information} \cite{Cover} between the system infinite state sequence $x^{\infty}= (x_1,...,x_{\infty})$ and its optimal estimate $\hat{x}^{\infty}= (\hat{x}_1,...,\hat{x}_{\infty})$ obtained by Kalman filtering given the infinite sequence of distorted disclosed data. Mutual information $I(x^{\infty};\hat{x}^{\infty})$ between the two jointly distributed infinite-dimensional vectors, $x^{\infty}$ and $\hat{x}^{\infty}$, is a measure of the statistical dependence between them. We design the privacy mechanisms to minimize $I(x^{\infty};\hat{x}^{\infty})$ for a desired maximum level of control performance degradation on the closed-loop infinite horizon LQR control cost. As we prove in this manuscript, we can cast the problem of finding sub-optimal additive random vectors covariance matrices and coordinate transformations as a constrained convex program (convex cost with LMI constraints). %%% improve
This is the first piece of work that provides privacy-preserving design tools for MIMO multidimensional feedback control systems to minimize infinite horizon bidirectional information flow by optimally distorting disclosed data while maintaining prescribed control performance. Providing infinite-horizon privacy is important in the context of dynamical systems since adversaries can infer information about private data from disclosed data over time. 
%An extended version of this paper containing additional results and discussions is available in --Arxiv--.
%%%%%%%%%%%%%%%%%%%%%%%%%%%%%%%%%%%%%%%%%%%%%%%%%%%%%%%%%%%%%%%%%%%
%%%%%%%%%%%%%%%%%%%%%%%%%%%%%%%%%%%%%%%%%%%%% from here
\section{Problem Formulation}
\subsection{System Description}
We consider the networked control architecture shown in Fig. 1. The dynamics of the plant is described as follows:
\begin{eqnarray}\label{eq1}
\mathcal{P} := \left\{ \begin{aligned}
x_{k+1} &= Ax_k + Bu_k+ w_k,\\
y_k &= x_k + h_k,\\
u_k&=K y_k,
\end{aligned} \right.
\end{eqnarray}
with time-index $k \in \Nat$, state $x_k \in {\mathbb{R}^{{n_x}}}$, measurable output $y_k \in {\mathbb{R}^{{n_y}}}$, controller $u_k \in {\mathbb{R}^{{n_u}}}$ with control feedback gain $K$, and matrices $(A,B,K)$ of appropriate dimensions, ${n_x}, {n_y}, {n_u} \in \mathbb{N}$. The state and output disturbances $w_k$ and $h_k$ are multivariate i.i.d. Gaussian processes with zero mean and covariance matrices ${\Sigma ^w}>0$ and ${\Sigma ^h}>0$, respectively. The initial state $x_1$ is assumed to be a Gaussian random vector with zero mean and covariance matrix $\Sigma^x_1 := E[x_1x_1^{\top}] $, $\Sigma^x_1 > 0$. Disturbances $w_k$ and $h_k$ and the initial condition $x_1$ are mutually independent. We assume that matrices $(A,B,\Sigma^x_1,\Sigma^w,\Sigma^h,K)$ are known, and $(A,B)$ is stabilizable.

We consider the setting where the local plant is controlled by a remote station. The user who owns the plant transmits $y_k$ to the remote station through an unsecured/public communication network to compute control actions (a remote LQR controller). Then, the control signal $u_k$ is sent back to the user through the network. To characterize control performance for some given positive definite matrices $Q$ and $R$, we introduce the associated infinite horizon LQR cost:
\begin{equation}\label{LQR_1}
C_\infty(x,u) := \limsup _{N \rightarrow \infty} \frac{1}{N+1} \sum_{k=0}^N \mathbb{E}\left({x_k^\top} Q x_k +{u_k^\top} R u_k\right),
\end{equation}
where $\mathbb{E}(\cdot)$ denotes expectation.

\begin{figure}[t]
  \includegraphics[width=3.5in]{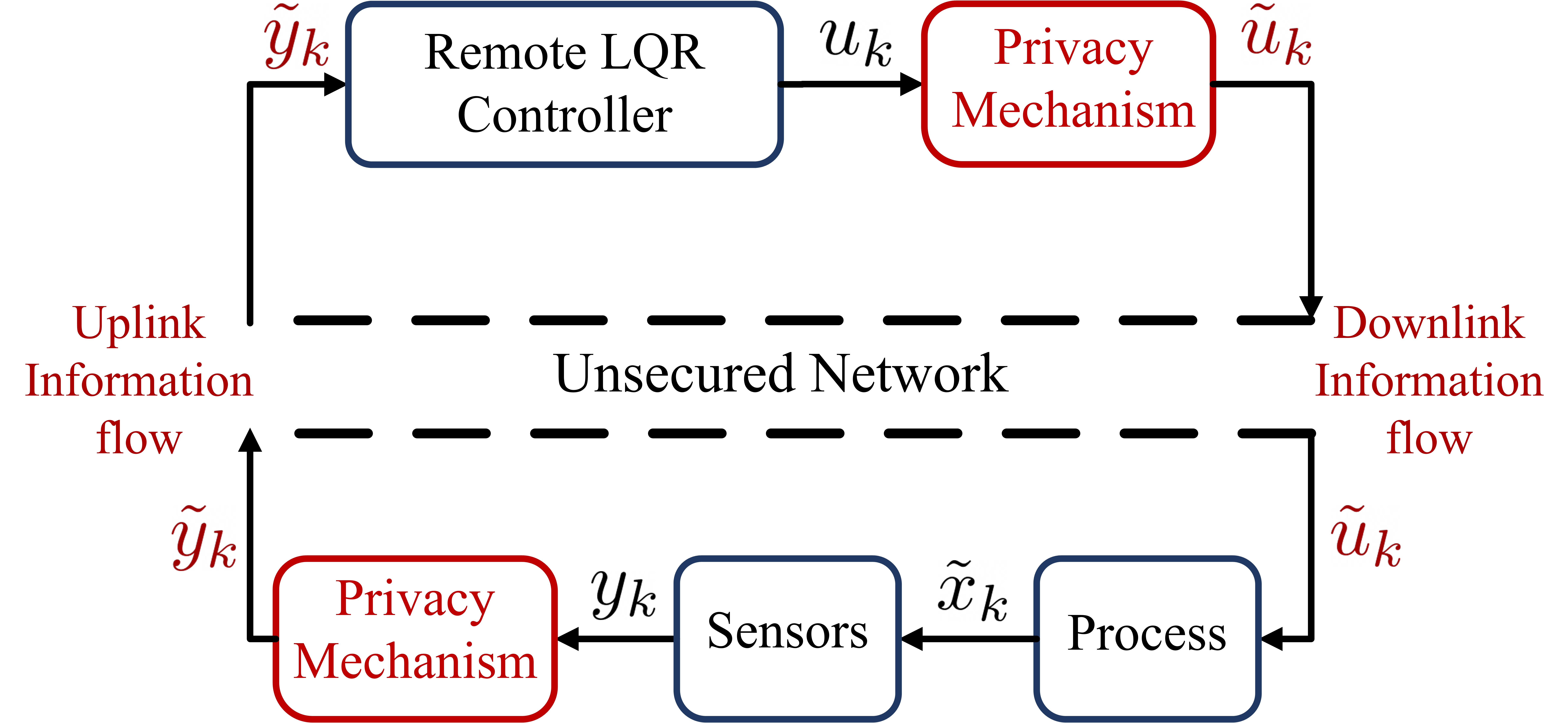}
    \caption{System configuration.}
    \label{fig2}
\end{figure}

For privacy reasons, a full disclosure of the state trajectory $x_k$, $k \in \mathbb{N}$ is not desired. We aim to prevent adversaries from estimating $x_k$ accurately. To this end, the user randomize measurement data $y_k$ before disclosure, and requests the remote station to randomize control signals, $u_k$, before transmission. By doing so, we protect against inference at the network and remote station. The idea is to distort $y_k$ and $u_k$ through random affine transformations of the form:
\begin{eqnarray}\label{eq2}
\mathcal{M} := \left\{ \begin{aligned}
\tilde{y}_k &= Gy_k + v_k,\\
\tilde{u}_k &= u_k + z_k,
\end{aligned} \right.
\end{eqnarray}
where $G \in {\mathbb{R}^{{n_y} \times {n_y}}}$ is a linear transformation, and $v_k$ and $z_k$ are zero mean i.i.d. Gaussian processes with covariance matrices $\Sigma^v$ and $\Sigma^z$, respectively. The distorted vectors $\tilde{y}_k$ and $\tilde{u}_k$ are transmitted over the network, see Fig. \ref{fig2}. It follows that the closed-loop dynamics when the privacy mechanism \eqref{eq2} is acting on the system is given by
\begin{eqnarray}\label{eq1priv}
\tilde{\mathcal{P}} := \left\{ \begin{aligned}
\tilde{x}_{k+1} &= A\tilde{x}_k + B\tilde{u}_k+ w_k,\\
\tilde{y}_k &= G\tilde{x}_k + Gh_k + v_k,\\
\tilde{u}_k&=KG\tilde{x}_k + KGh_k + Kv_k + z_k.
\end{aligned} \right.
\end{eqnarray}
with distorted state $\tilde{x} \in \mathbb{R}^{n_x}$. Here, we seek to synthesize $G$, $\Sigma^v$, and $\Sigma^z$, to make estimating the infinite horizon state trajectory $\tilde{x}_k$, $k \in \mathbb{N}$, as ``hard'' as possible from the disclosed data, $(\tilde{y}_k,\tilde{u}_k)$, $k \in \mathbb{N}$.

%%%%%%%%%%%%%%%%%%%%%%%%%%%%%%%%%%%%%%%%%%kalman filter

We assume the adversary uses a steady-state Kalman filter designed to estimate the state \emph{in the absence of privacy mechanisms}. That is, we assume the adversary has prior knowledge of the system dynamics (matrices ($A,B,\Sigma^x_1,\Sigma^w,\Sigma^h$)
but does not have knowledge about the privacy mechanism (matrices ($G,\Sigma^v,\Sigma^z$). This creates an asymmetry we seek to exploit to increase privacy. The considered filter has the following structure:
\begin{equation}\label{kalmanfilter}
\left\{
    \begin{aligned}
    &\hat{x}_{k|k-1} = A\hat{x}_{k-1}+Bu_{k-1},\\
    &\hat{x}_k = \hat{x}_{k \mid k-1}+L\left(\tilde{y}_k- \hat{x}_{k \mid k-1}\right),
\end{aligned}       
\right.
\end{equation}
with estimated state $\hat{x}_k \in \mathbb{R}^{n_x}$ and gain  $L \in \mathbb{R}^{n_x \times n_y}$. The adversary designs the filter for the distortion-free system \eqref{eq1}. Let $\rho_k$ denote the estimation error \emph{in the absence of the privacy distortions} $\rho_k := x_k - \hat{x}_k$. The observer gain $L$ is designed to minimize the asymptotic covariance matrix $\Sigma^\rho := \lim_{k \rightarrow \infty} E\left(\rho_k \rho_k^{\top}\right)$ \cite{Astrom1997}. Because the system is observable (we have state measurements), $\Sigma^\rho$ always exists. 

Now let $e_k$ denote the estimation error \emph{in the presence of privacy distortions}, i.e., $e_k := \tilde{x}_k - \hat{x}_k$. Given the distorted dynamics \eqref{eq1priv}, the privacy mechanisms \eqref{eq2}, and the estimator \eqref{kalmanfilter}, the estimation error dynamics is governed by the following coupled difference equations:
\begin{equation}\label{estimationerrordynamics}
\left\{
\begin{aligned}
\tilde{x}_{k+1} &=(A+B K G) \tilde{x}_k+B K \tilde{v}_k+B z_k+w_k,\\
e_{k|k-1}&=Ae_{k-1}+Bz_{k-1}+w_{k-1},\\
e_k &=(I-L) e_{k \mid k-1}-L(G-I) \tilde{x}_k-L \tilde{v}_k,
\end{aligned}
\right.
\end{equation}
%%%%%%%%%%%%%%%%%%%%%%%%%
where $\tilde{v}_k := Gh_k+v_k$. %In what follows, we introduce the adversarial model we seek to defend against.
%\subsection{Adversarial Capabilities}
%We consider worst-case adversaries that eavesdrop data at the communication network and/or the remote station. They do not only have access to all distorted sensor measurements $\tilde{y}_k$ and distorted input signal $\tilde{u}_k$, but also have prior knowledge of the dynamics and the stochastic properties of the system, i.e., matrices  $(A,B,K,\Sigma^x_1,\Sigma^w,\Sigma^h)$ are known by the adversary. Moreover, the adversary also knows covariance matrices $(\Sigma^{\tilde{y}}_k,\Sigma^{\tilde{u}}_k)$ as these can be estimated from the disclosed data $(\tilde{y}_k,\tilde{u}_k)$. In practice, actual adversaries would typically not have all the capabilities that we assume here. However, if we maximize privacy under such worst-case adversaries, we ensure that adversaries with less capabilities perform even worse.
%%%%%%%%%%%%%%%%%%%%%%%%%%%%%%%%%%%%%%%
%Privacy and Performance Metrics and  ??

\subsection{Problem Formulation}

The aim of our privacy scheme is to make the estimation of the infinite horizon state sequence, ${\tilde{x}^{\infty}} := (\tilde{x}_1,\ldots,\tilde{x}_{\infty})$, from the disclosed distorted data, ${\tilde{y}^{\infty}} := (\tilde{y}_1,\ldots,\tilde{y}_{\infty})$ and \linebreak${\tilde{u}^{\infty}} := (\tilde{u}_1,\ldots,\tilde{u}_{\infty})$, as hard as possible without degrading the control performance excessively. Hence, when designing the distorting variables $(G,\Sigma^v,\Sigma^h)$, we need to consider the \emph{trade-off between privacy and performance}.

As privacy metric, we use the mutual information rate $I_\infty(\tilde{x};\hat{x})$ \cite{Cover} between $\tilde{x}^{\infty}$ and the infinite sequence of estimates $\hat{x}^{\infty} := (\hat{x}_1,...,\hat{x}_{\infty})$
obtained by Kalman filtering:
\begin{equation}\label{MI_1}
I_\infty(\tilde{x};\hat{x}) := \limsup _{N \rightarrow \infty} \frac{1}{N+1} I(\tilde{x}^N;\hat{x}^N),
\end{equation}
where $I(\tilde{x}^N;\hat{x}^N)$ denotes standard mutual information \cite{Cover}.

We use the LQR cost in \eqref{LQR_1} to quantify control performance \emph{in the absence of attacks}. To quantify the effect of the privacy mechanism \eqref{eq2} on the control performance, we introduced the associated distorted LQR control cost: 
\begin{equation}\label{LQR_2}
\tilde{C}_\infty(\tilde{x},\tilde{u}) := \limsup _{N \rightarrow \infty} \frac{1}{N+1} \sum_{k=0}^N \mathbb{E}\left({\tilde{x}_k^\top} Q \tilde{x}_k +{\tilde{u}_k^\top} R \tilde{u}_k\right).
\end{equation}

We aim to minimize $I_{\infty}(\tilde{x};\hat{x})$ subject to a constraint on the LQR cost increase due to the privacy mechanism, $\tilde{C}_{\infty}(\tilde{x},\tilde{u}) - C_{ \infty}(x,u) \le \epsilon$, for a desired maximum control performance degradation level ${\epsilon} \in {\mathbb{R}^+}$, using as synthesis variables the mechanism matrices $G$, ${\Sigma^v}$, and ${\Sigma^z}$. In what follows, we present the problem we seek to address.

\vspace{1mm}
\noindent
\textbf{Problem 1} Given the system dynamics \eqref{eq1}, distortion-free control performance \eqref{LQR_1}, distorted control performance \eqref{LQR_2}, privacy mechanism \eqref{eq2}, distorted dynamics \eqref{eq1priv}, Kalman filter \eqref{kalmanfilter}, and maximum control degradation level ${\epsilon}>0$, find the privacy mechanism variables, $G$, ${\Sigma^v}$, and ${\Sigma^z}$, solution of the following optimization problem:
\begin{equation} \label{problem1}
\left\{\begin{aligned}
    &\min_{G, {\Sigma^v}, {\Sigma^z}} I_{\infty}(\tilde{x} ; \hat{x}),\\[1mm]
    &\hspace{4mm}\text{s.t. } \tilde{C}_{\infty}(\tilde{x},\tilde{u}) - C_{ \infty}(x, u) \le \epsilon.
\end{aligned}\right.
\end{equation}
%Solution to Problem 1
\section{Privacy Mechanism Design}
To solve Problem 1, we first need to write the cost function and constraint in terms of the design variables.
\subsection{Cost Function: Formulation and Convexity}
Mutual information $I\left(\tilde{x}^N; \hat{x}^N\right)$ can be written in terms of uplink
$I\left(\tilde{x}^N \rightarrow \hat{x}^N\right)$ (plant to the remote station) and downlink $I\left(\tilde{x}^N \leftarrow \hat{x}^N\right)$ (remote station to the plant) directed information flows \cite{massey2005conservation}:
\begin{equation}
I\left(\tilde{x}^N ; \hat{x}^N\right)=I\left(\tilde{x}^N \rightarrow \hat{x}^N\right)+I\left(\tilde{x}^N \leftarrow \hat{x}^N\right) .
\end{equation}
Then, the mutual information rate can be written as
\begin{align}
    &I_{\infty}(\tilde{x} ; \hat{x}) :=  \limsup _{N \rightarrow \infty} \frac{1}{N+1} \left(I\left(\tilde{x}^N \rightarrow \hat{x}^N\right)\right. \nonumber\\
    &\hspace{45mm} +\left.I\left(\tilde{x}^N \leftarrow \hat{x}^N\right)\right).\label{infinitemutualinf}
\end{align}
The decomposition of $I\left(\tilde{x}^N ; \hat{x}^N\right)$ in terms of uplink and downlink directed information is essential in enabling us to express mutual information as a stage additive function of covariance matrices. The latter allows writing $I_{\infty}(\tilde{x}; \hat{x})$ in terms of the solution of Lyapunov equations/inequalities, which in turn enables a convex reformulation of cost and constraints. In Lemma 1, we write the resulting expression of $I\left(\tilde{x}^N ; \hat{x}^N\right)$ in terms of the design variables. Then, $I_{\infty}(\tilde{x}; \hat{x})$ can be obtained by taking the limit in \eqref{infinitemutualinf}. Please refer to the proof of Lemma 1 for a step by step derivation of $I\left(\tilde{x}^N ; \hat{x}^N\right)$.
%%%%%%%%%%%%%%%%%%%%%%%%%% Lemma 1
\begin{lemma}
Mutual information $I\left(\tilde{x}^N ; \hat{x}^N\right)$ can be written in terms of $G$, ${\Sigma^v}$, and ${\Sigma^z}$, as follows:
\begin{equation}
\begin{aligned}
&I\left(\tilde{x}^N ; \hat{x}^N\right)=\sum_{k=1}^N\left(\frac{1}{2} \log \operatorname{det}\left(LG \Sigma_{k|k-1}^e G^\top L^\top +L \Sigma^{\Tilde{v}} L^{\top}\right)\right.\\
&\qquad-  \frac{1}{2} \log \operatorname{det}\left(L\Sigma^{\tilde{v}} L^{\top}\right)-\frac{1}{2} \log \operatorname{det}\left(B \Sigma^zB^\top + \Sigma^w\right)\\
&\qquad+\left.\frac{1}{2} \log \operatorname{det}\left(BK \Sigma^{\tilde{v}}K^{\top}B^\top +B \Sigma^zB^\top + \Sigma^w\right)\right),
\end{aligned}\label{lemma1}
\end{equation}
with covariance matrices $\Sigma_{k|k-1}^e := \mathbb{E}(e_{k|k-1} e_{k|k-1}^\top)$ and $\Sigma^{\tilde{v}} := G \Sigma^h G^\top + \Sigma^v$.
\end{lemma}
\textbf{\emph{Proof}}:
See  Appendix A.
\hfill $\blacksquare$\\

%%%%%%%%%%%%%%% why SigmaVt
\vspace{-2mm}
Note that $\Sigma^v$ only appears in the expression for $\Sigma^{\tilde{v}}$. Given $(G,\Sigma^{\tilde{v}})$, matrix $\Sigma^v$ is fully determined and vice versa. That is, $(G,\Sigma^{\tilde{v}})\rightarrow (G,\Sigma^{v})$ is an invertible transformation. Therefore, we can pose both the cost and constraint of Problem 1 in terms of either $\Sigma^{\tilde{v}}$ or $\Sigma^{{v}}$. Casting the problem in terms of $\Sigma^{\tilde{v}}$ allows us to write convex cost and constraint. Hereafter, we pose the problem in terms of $(G,\Sigma^{\tilde{v}})$. Once we have found optimal $(G,\Sigma^{\tilde{v}})$, we extract the optimal $\Sigma^v$ as $\Sigma^v = \Sigma^{\tilde{v}} - G \Sigma^h G^\top$. Note, however, that due to the negative term $-G\Sigma^h G^\top$, the extracted $\Sigma^v$ might be negative semidefinite, which is of course wrong as $\Sigma^v$ is a covariance matrix. To avoid this, we enforce that the extracted $\Sigma^v$ is always positive definite in the synthesis program by adding $\Sigma^{\tilde{v}} -G \Sigma^h G^\top  > \mathbf{0}$ as an extra constraint. This constraint can be equivalently written as the following linear inequality in $(G,\Sigma^{\tilde{v}})$ using Schur complement properties \cite{zhang2006schur}: %properties \cite{zhang2006schur}, the later nonlinear inequality can be rewritten as a LMI in $\Sigma^{\tilde{v}}$ and $G$ as follows:
\begin{eqnarray}\label{SigmaV_pos_def}
\left[ {\begin{array}{*{20}{c}}
\Sigma^{\tilde{v}}    &   G \\
{G}^\top      &    (\Sigma^h)^{-1}
\end{array}} \right] > \mathbf{0}.
\end{eqnarray}
We use inequality \eqref{SigmaV_pos_def} later when we solve the complete optimization problem to enforce that the optimal $(G,\Sigma^{\tilde{v}})$ leads to a positive definite $\Sigma^v$.
%%%%%%%%%%%%%%%%%%%%%%%%%%%%%%%%%Covariance of estimation error

In Lemma 1, we have an expression of mutual information in terms of the design variables and the estimation error covariance $\Sigma_{k|k-1}^e$. Consider the closed-loop dynamics \eqref{estimationerrordynamics}, and define the extended state  $\zeta_k := \text{col}\left[ e_{k \mid k-1},  \tilde{x}_k
\right]$, we have
\begin{equation}
\begin{aligned}
\zeta_{k+1}&=\left[\begin{array}{cc}
A(I-L)& -AL(G-I) \\
\mathbf{0} & A+BK G
\end{array}\right] \zeta_k \\ &+\left[\begin{array}{ccc}
-AL &  B & I \\
BK  & B & I
\end{array}\right]\left[\begin{array}{c}
\tilde{v}_k\\
z_k\\
w_k
\end{array}\right].
\end{aligned}
\end{equation}
Because $(\tilde{v}_k,z_k,w_k)$ are all zero mean i.i.d. processes, the covariance of $\zeta_k$, $\Sigma^\zeta_k := \mathbb{E}(\zeta_k\zeta_k^\top)$, satisfies the following:
\begin{equation}\label{Riccatiricursion}
\begin{aligned}
\Sigma_{k+1}^\zeta&=\mathcal{A} \Sigma_k^\zeta\mathcal{A}^T+\mathcal{B},
\end{aligned}
\end{equation}
where
\begin{equation}\label{mathcalAB}
\left\{\begin{aligned}
&\mathcal{A}:=\left[\begin{array}{cc}
A(I-L)& -AL(G-I) \\
\mathbf{0} & A+BK G
\end{array}\right],\\
&\mathcal{B}:= \begin{bmatrix}
-AL & \hspace{-1.75mm}B & \hspace{-1.75mm}I\\ BK & \hspace{-1.75mm} B &\hspace{-1.75mm} I
\end{bmatrix} \begin{bmatrix}
\Sigma^{\tilde{v}} & \hspace{-2mm} &  \\  &  \hspace{-2mm} \Sigma^z & \\& &  \hspace{-2mm} \Sigma^w
\end{bmatrix}\begin{bmatrix}
-AL & \hspace{-1.75mm}B & \hspace{-1.75mm}I\\ BK & \hspace{-1.75mm} B &\hspace{-1.75mm} I
\end{bmatrix}^\top.
%&\mathcal{B}:=\begin{bmatrix}
%-AL & \hspace{-2.5mm} BK \\ B & \hspace{-2.5mm} B \\ I & \hspace{-2.5mm} I
%\end{bmatrix}^\top
%\begin{bmatrix}
%\Sigma^{\tilde{v}} & \hspace{-2.5mm} &  \\  &  \hspace{-2.5mm} \Sigma^z & \\& &  \hspace{-2.5mm} \Sigma^w
%\end{bmatrix}
%\begin{bmatrix}
%-AL & \hspace{-2.5mm} BK \\ B & \hspace{-2.5mm} B \\ I & \hspace{-2.5mm} I
%\end{bmatrix}.
\end{aligned}\right.
\end{equation}
If $\mathcal{A}$ is Schur stable (which is always the case for $G=I$ by construction), the limit $\Sigma^\zeta := \lim _{k \rightarrow \infty} \Sigma^{\zeta}_k$, with $\Sigma^{\zeta}_k$ solution of \eqref{Riccatiricursion}, exists and coincides with the unique positive definite solution of the Lyapunov equation:
\begin{equation}\label{lyapanov}
    \mathcal{A} \Sigma^\zeta \mathcal{A}^T-\Sigma^\zeta + \mathcal{B}=\mathbf{0}.
\end{equation}
Moreover, because $\zeta_k = \text{col}\left[ e_{k \mid k-1},  \tilde{x}_k
\right]$, we have
\begin{align}\label{Sigmax}
\Sigma^{\tilde{x}} &:= \lim _{k \rightarrow \infty} \Sigma_{k}^{\tilde{x}} = \begin{bmatrix} \mathbf{0} &  I \end{bmatrix} \Sigma^\zeta \begin{bmatrix} \mathbf{0} &  I \end{bmatrix}^\top,\\
\Sigma^e &:= \lim _{k \rightarrow \infty} \Sigma_{k|k-1}^e = \begin{bmatrix} I & \mathbf{0} \end{bmatrix} \Sigma^\zeta \begin{bmatrix} I & \mathbf{0} \end{bmatrix}^\top, \label{Sigmae}
\end{align}
which allows writing the following corollary of Lemma 1 by taking the limit in \eqref{infinitemutualinf}.

\vspace{-1mm}

\begin{corollary}\label{corollary_1}
The mutual information rate $I_{\infty} \left(\tilde{x}; \hat{x}\right)$ in \eqref{lemma1} can be written in terms of $G$, ${\Sigma^{\tilde{v}}}$, and ${\Sigma^z}$, as follows:
\begin{equation}
\begin{aligned}
&I_\infty\left(\tilde{x}; \hat{x}\right) = \frac{1}{2} \log \operatorname{det}\left(LG \Sigma^e G^\top L^\top +L \Sigma^{\Tilde{v}} L^{\top}\right)\\
&\quad-  \frac{1}{2} \log \operatorname{det}\left(L\Sigma^{\tilde{v}} L^{\top}\right)-\frac{1}{2} \log \operatorname{det}\left(B \Sigma^zB^\top + \Sigma^w\right)\\
&\quad+\frac{1}{2} \log \operatorname{det}\left(BK \Sigma^{\tilde{v}}K^{\top}B^\top +B \Sigma^zB^\top + \Sigma^w\right),
\end{aligned} \label{cost_a}
\end{equation}
with $\Sigma^e = \lim _{k \rightarrow \infty} \Sigma_{k|k-1}^e$ as defined in \eqref{Sigmae}.
\end{corollary}
Note that the cost $I_\infty\left(\tilde{x}; \hat{x}\right)$ in \eqref{cost_a} is non-convex in the design variables. The term $LG\Sigma^eG^\top L^\top$ is quadratic in $G$ and $\Sigma^e$ depends on the solution of the Lyapunov equation \eqref{lyapanov}, which is itself quadratic in $G$. To tackle this, we derive a convex upper bound on the cost \eqref{cost_a} and minimize this bound. We start with an upper bound, $\Sigma$, on the solution $\Sigma^\zeta$ of the Lyapunov equation \eqref{lyapanov}. Having this $\Sigma$ and using \eqref{Sigmae} and monotonicity of $\log \operatorname{det} (\cdot)$ allow us to upper bound the first term of the cost in \eqref{cost_a}. In the following lemma, we propose a convex program to find $\Sigma$.
%%%%%%%%%%%%%%%%%%%%%%%%%%%%%%%%%%%%%%%%%%%%%%%%%%%%%%%%%%%%%%%%%%%%%%%%%%%%%%%%%%%%%%%%%% Lemma 3
\begin{lemma}\label{mutualinformationcov}
An upper bound $\Sigma$ on the solution $\Sigma^\zeta$ of \eqref{lyapanov} can be found by solving the following convex program:
%%%%%%%%%%%%%
\begin{equation}\label{pr0}
\left\{\begin{aligned}
&\min_{{\Sigma},\Pi_1,\Pi_2} \operatorname{trace}(\Sigma),\\[1mm]
&\hspace{4mm}\text{s.t. }
\left[\begin{array}{cc}
\Sigma -{\mathcal{B}} & \mathcal{A}_{0} \Pi_1 + \mathcal{A}_{1} \Pi_2 \\
* & \Pi_1 +\Pi_1^{\top}- \Sigma
\end{array}\right] \ge \mathbf{0},\\
&\hspace{11mm} \Pi_1 =\left[\begin{array}{cc}
 \Pi_{11}&\Pi_{12}\\
 \mathbf{0}&\Pi_{13}
\end{array}\right],\\
&\hspace{11mm}\Pi_2 = \left[\begin{array}{cc}
\mathbf{0} &\Pi_{21}
\end{array}\right],
\end{aligned}\right.
\end{equation}
where
\begin{equation}\label{definematrices}
\left\{\begin{aligned}
&{\mathcal{A}}_{0}:= \left[\begin{array}{cc}
A(I-L) &AL \\
\mathbf{0} & A
\end{array}\right],\,\, {\mathcal{A}}_{1}:= \left[\begin{array}{c}
-AL \\
BK 
\end{array}\right].
\end{aligned}\right.
\end{equation}
\emph{\textbf{\emph{Proof}}:
See  Appendix B.
\hfill $\blacksquare$}
\end{lemma}

%%%%%%%%%%%%%%%%%%%%%%%%%%%%%%%%%%%%%%%%%%%%%%%%%%%%%%%%%%%%%%%%%%
%%%%%%%%%%%% solve in terms of pi5 rather than G
We defined new variables $\Pi_1$ and $\Pi_2$ to convexity the constraints in \eqref{pr0}. Given $(\Pi_{1},\Pi_{21})$, matrix $G$ can be extracted as $G = \Pi_{21}\Pi_{13}^{-1}$ (see the proof of Lemma 2). Therefore, we can pose both cost and constraints in terms of either $G$ or $\Pi_{21}$. Casting the problem in terms of $\Pi_{21}$ allows us to linearize some constraints. Hereafter, we pose the problem in terms of $(\Pi_1,\Pi_{21})$. Once we have found optimal $(\Pi_1,\Pi_{21})$, we extract the optimal $G$ using $\Pi_{21}= G \Pi_{13}$.

Lemma 2 allows casting the computation of an upper bound, $\Sigma$, on the solution, $\Sigma^{\zeta}$, of the Lyapunov equation \eqref{lyapanov} as the solution of an optimization problem. Matrix $\Sigma$ obtained by solving \eqref{pr0} satisfies $\Sigma \geq \Sigma^{\zeta} = \lim _{k \rightarrow \infty} \Sigma^{\zeta}_k$. Therefore, given $\Sigma$, by \eqref{Sigmax}-\eqref{Sigmae}, we also have the following upper bounds on $\Sigma^{\tilde{x}}$ and $\Sigma^e$
\begin{equation}\label{estimationerrorcov}
 \left\{\begin{aligned}
 &\Sigma^{\tilde{x}} = \lim _{k \rightarrow \infty} \Sigma^{\tilde{x}}_{k} \leq N_{\tilde{x}} \Sigma N_{\tilde{x}}^\top,\\
&\Sigma^e = \lim _{k \rightarrow \infty} \Sigma^{e}_{k|k-1} \leq  N_{e} \Sigma N_{e}^\top,\\
&N_{\tilde{x}} := \begin{bmatrix} \mathbf{0} &  I \end{bmatrix},
N_{e} := \begin{bmatrix} I &  \mathbf{0} \end{bmatrix}. \\
\end{aligned}\right.
\end{equation}
In Corollary  1, the mutual information rate is written in terms of privacy mechanism variables and $\Sigma^e$. Hence, given \eqref{estimationerrorcov} and monotonicity of the determinant function, an upper bound on $I_{\infty}(\tilde{x};\hat{x})$ in terms of $\Sigma$ can be written as follows:
%Given Lemma 1 and \eqref{estimationerrorcov}, the mutual information can be expressed as follows:
 \begin{equation}\label{costf}
 \left\{\begin{aligned}
&I_{\infty}(\tilde{x} ; \hat{x})\le\frac{1}{2} \log \operatorname{det}\left(LG N_{e} \Sigma N_{e}^\top G^{\top}L^\top +L \Sigma^{\tilde{v}}L^{\top}\right)\\
&\qquad -  \frac{1}{2} \log \operatorname{det}\left(L\Sigma^{\tilde{v}} L^{\top}\right)- \frac{1}{2} \log \operatorname{det}\left(B \Sigma^zB^\top + \Sigma^w\right)\\
&\qquad+\frac{1}{2} \log \operatorname{det}\left(BK_{c} \Sigma^{\tilde{v}}K_{c}^{\top} B^\top +B \Sigma^zB^\top + \Sigma^w\right).
\end{aligned}\right.
\end{equation}
%%%%%%%%%%%%%%%%%%%%%%%%%%%%%%%%%%%%%%%%%

So far, we have an upper bound \eqref{costf} on the cost function in Problem 1 in terms of the solution $\Sigma$ of program \eqref{pr0} and the mechanism parameters. However, \eqref{costf} is still non-convex in $G$ and $\Sigma$. In Lemma 3, we pose the problem of minimizing the right-hand side of \eqref{costf} as a convex program. This reformulation is achieved using Schur complement properties, an epigraph reformulation of the minimization problem, and the monotonicity of the logdet$(\cdot)$ function. Moreover, as we will later need to combine the program in Lemma 2 with the convex reformulation of the bound in \eqref{costf}, we write, in Lemma 3, $G$ in terms of $\Pi_{2}$ and $\Pi_{1}$ as we do in Lemma 2 ($G = \Pi_{21}\Pi_{13}^{-1}$, see the discussion below Lemma 2). This is necessary as we have to use the same coordinates in the reformulation of cost and constraints to be able to later solve all together as a single optimization problem.

%%%%Lemma 2
%\textbf{\emph{Lemma 2}}: 
\begin{lemma}\label{mutualinformationconvex}
Consider the solution of the convex program: 
\begin{equation}\label{inequalityofcost1} 
\left\{\begin{aligned}
&\min _{\Pi_{13},\Pi_{21},\Pi_3,\Pi_4,\Sigma^{\tilde{v}},\Sigma^z,\Sigma}\left( -\frac{1}{2} \text{\emph{logdet}} (\Pi_3) -\frac{1}{2} \text{\emph{logdet}} \left(\Pi_4\right) \right.\\&\left. - \frac{1}{2} \text{\emph{logdet}} (L\Sigma^{\tilde{v}} L^{\top}) - \frac{1}{2} \text{\emph{logdet}} \left(B \Sigma^zB^\top + \Sigma^w\right)\right)\\[1mm]
& \left\{\begin{aligned} &\text{\emph{s.t. }}\\
&\left[ {\begin{array}{*{20}{c}}
2I-\Pi_3-L \Sigma^{\tilde{v}}L^\top & L \Pi_{21}\\
*& \Pi_{13} +\Pi_{13}^{\top} -  N_e \Sigma N_e^\top
\end{array}} \right] \ge \mathbf{0},\\
&2I-\Pi_4 \ge \left(BK \Sigma^{\Tilde{v}}K^\top B^\top +B \Sigma^zB^\top + \Sigma^w \right).
\end{aligned}\right.
\end{aligned}\right.
\end{equation}
The resulting $\Sigma$, $\Sigma^{\tilde{v}}$, $\Sigma^z$, and $G = \Pi_{21}\Pi_{13}^{-1}$ minimize the upper bound on $I_{\infty}(\tilde{x} ; \hat{x})$ in \eqref{costf}. 
\end{lemma}
\textbf{\emph{Proof}}:
See  Appendix C.
\hfill $\blacksquare$
%

%%%%%%%%%%%%%%

By Lemma $1$, Lemma $2$, and Lemma $3$, a minimal upper bound on the cost $I_{\infty}(\tilde{x} ; \hat{x})$ can be achieved by solving the convex programs in \eqref{pr0} and \eqref{inequalityofcost1}. Then, if the constraints on positive definiteness of $\Sigma^v$ \eqref{SigmaV_pos_def} and control performance, $\tilde{C}_{\infty}(\tilde{x},\tilde{u}) - C_{ \infty}(x,u) \le \epsilon$, can be written as convex functions of the decision variables, we can find optimal distorting mechanisms efficiently using off-the-shelf optimization algorithms. Regarding \eqref{SigmaV_pos_def}, it can be verified (see Appendix D) that \eqref{SigmaV_pos_def} can be written in terms of $(\Pi_{13},\Pi_{21})$, the new decision variables, instead of the original $G$, as follows:
\begin{eqnarray}\label{SigmaV_pos_def2}
\left[ {\begin{array}{*{20}{c}}
\Sigma^{\tilde{v}}    &   \Pi_{21} \\
*    &    \Pi_{13}+\Pi_{13}^\top-\Sigma^h
\end{array}} \right] \ge \mathbf{0}.
\end{eqnarray}
We will add this \eqref{SigmaV_pos_def2} as a new constraint in the synthesis program. It remains to reformulate the control constraint.

\subsection{Control Performance: Formulation and Convexity}

\begin{lemma}\label{constraintlemma}
%\textbf{\emph{Lemma 4}}:\\
The constraint on the LQR control cost:
\begin{equation}\label{constraint}
    \tilde{C}_{\infty}(\tilde{x}, \tilde{u}) -  {C}_{\infty}(x, {u}) \le \epsilon,
\end{equation}
%where
%\begin{equation}
%    \tilde{C}_{\infty}(x, \tilde{u}) \triangleq \limsup _{K \rightarrow \infty} \frac{1}{K+1} \sum_{k=0}^N \mathbb{E}\left[x_k^\top Q x_k + \tilde{u}_k^\top R \tilde{u}_k\right],
%\end{equation}
can be formulated as the following set of LMIs:
\begin{equation}\label{pr30}
\left\{\begin{aligned}
&\text {tr}\left( Q \Sigma^{\tilde{x}}\right)+\text {tr}\left(\Pi_5\right)\\
&\qquad+\operatorname{tr}\left(K^{\top} R K \Sigma^{\tilde{v}}+R\Sigma^z\right) \le C_{\infty}(x, u)+\epsilon,\\
& \left[\begin{array}{cc}
\Pi_5 & R^{1/2} K \Pi_{21} \\
* & \Pi_{13} +\Pi_{13}^{\top} - \Sigma^{\tilde{x}}
\end{array}\right] \ge \mathbf{0},
\end{aligned}\right.
\end{equation}
with new matrix variable $\Pi_5$ to be designed.
%%%%%%%%%%%%%%%%%
%%%%%%%%%%%%%%%%%
\end{lemma}
\textbf{\emph{Proof}}:
See  Appendix E.
\hfill $\blacksquare$\\
%%%%%%%%%%%%%%%%%%%%%%%%%%%%%%%%%%%%%%%%%%%%%%%%%%%%%%%
%%%%%%%%%%%%%%%%%%%%%%%\\

In Lemma 1 - Lemma 4, an upper bound on the cost function $I_{\infty}(\tilde{x};\hat{x})$ and the distortion constraint $\tilde{C}_{\infty}(x, \tilde{u}) -  {C}_{\infty}(x, {u}) \le \epsilon$ are written in terms of convex functions (programs) of the design variables. We have, however, two cost functions in Lemma 2 and Lemma 3. The latter leads to a multi-objective optimization problem that can be solved by scalarizing the costs, i.e., introducing a single objective that represents a compromise between both of them. To this aim, we introduce $\alpha \in \mathbb{R}$, $\alpha>0$, as a weighting parameter and define a new cost as the weighted sum of the original ones (see the cost in \eqref{eq:convex_optimization15}). Since our goal is to achieve a minimal mutual information rate, because it characterizes information leakage, we seek the $\alpha$ that minimizes $I_{\infty}(\tilde{x};\hat{x})$ by performing a line search over $\alpha$ subject to all constraints in Lemma 1 - Lemma 4. In what follows, we pose the complete nonlinear convex program to find a sub-optimal solution for Problem 1 (sub-optimal in the sense that Lemma 3 seeks to minimize an upper bound on the actual cost).

%%%%%%%%%%%%%%%%%%%%%%%%%%%%
%%%%%%%%%%%%
\begin{table}
\noindent\rule{\hsize}{1pt}
\begin{equation} \label{eq:convex_optimization15}
\left\{\begin{aligned} 
&\min _{\Pi_1,\Pi_2,\Pi_3,\Pi_4,\Pi_5,\Sigma^{\tilde{v}},\Sigma^z,\Sigma} \alpha ( -\frac{1}{2} \log \operatorname{det} (\Pi_3)\\
& \frac{1}{2} \log \operatorname{det}(L\Sigma^{\tilde{v}} L^{\top})-\frac{1}{2} \log \operatorname{det}(\Pi_4)\\
&- \frac{1}{2} \log \operatorname{det}(B \Sigma^zB^\top + \Sigma^w)) +(1-\alpha) \operatorname{trace}(\Sigma),\\[1mm]
&\left\{\begin{aligned}
&\left[ {\begin{array}{*{20}{c}}
2I-\Pi_3-L \Sigma^{\tilde{v}} L^\top & \hspace{-1mm} L \Pi_{21}\\
* & \hspace{-1mm} \Pi_{13} +\Pi^{\top}_{13} - N_e \Sigma N_e^\top
\end{array}} \right] \ge \mathbf{0},\\
&2I-\Pi_4 \ge \left(BK \Sigma^{\Tilde{v}}K^\top B^\top +B \Sigma^z B^\top + \Sigma^w \right),\\
&{\left[\begin{array}{cc}
\Sigma -{\mathcal{B}} & \mathcal{A}_{0} \Pi_1 + \mathcal{A}_{1} \Pi_2 \\
*& \Pi_1 +\Pi_1^{\top} - \Sigma
\end{array}\right] \ge \mathbf{0}},\\
&\text {tr}\left( Q \Sigma^{\tilde{x}} \right)+\text {tr}\left(\Pi_5\right)\\
&\qquad \qquad +\text{tr}\left(K^{\top} R K \Sigma^{\tilde{v}} + R\Sigma^z\right)\le C_{\infty}(x, u)+\epsilon,\\
&{ \left[\begin{array}{cc}
\Pi_5 & R^{1/2} K \Pi_{21} \\
* & \Pi_{13} +\Pi^{\top}_{13} - N_{\tilde{x}} \Sigma N_{\tilde{x}}^\top
\end{array}\right] \ge \mathbf{0}},\\
&{\left[ {\begin{array}{*{20}{c}}
\Sigma^{\tilde{v}}    &   \Pi_{21}  \\
* & \Pi_{13} +\Pi^{\top}_{13}-\Sigma^h
\end{array}} \right]> \mathbf{0}, \,\,\, \Sigma^z > \mathbf{0}, \,\,\, \Sigma > \mathbf{0}.}
\end{aligned}\right.
\end{aligned}\right.
\end{equation}
\noindent\rule{\hsize}{1pt}
\vspace{-10mm}
\end{table}
%%%%%%%%%%%%%%%%%%%%%%%%%%%%%
%%%%%%%%%%%%%%%%%%%%%%%%%%%%%
\begin{theorem}\label{th1}
Consider the system dynamics \eqref{eq1}, distortion-free control performance \eqref{LQR_1}, distorted control performance \eqref{LQR_2}, privacy mechanism \eqref{eq2}, distorted dynamics \eqref{eq1priv}, Kalman filter \eqref{kalmanfilter}, and maximum control degradation level ${\epsilon}>0$, and matrices in \eqref{mathcalAB}, \eqref{definematrices}, and \eqref{estimationerrorcov}. For a fixed $\alpha>0$, given the solution of the convex program in \eqref{eq:convex_optimization15}, the mechanism variables $G$, $\Sigma ^v$, and $\Sigma^z$, that minimize the upper bound on $I_{\infty}(\tilde{x};\hat{x})$ in \eqref{costf} subject to the control performance degradation constraint, $\tilde{C}_{\infty}(x,\tilde{u}) - C_{ \infty}(x, u) \le \epsilon$, are given by $\Sigma^z$, $G = \Pi_{21}\Pi_{13}^{-1}$, and $\Sigma^v = \Sigma^{\tilde{v}} - \Pi_{21}\Pi_{13}^{-1} \Sigma^h (\Pi_{21}\Pi_{13}^{-1})^\top$. 
\end{theorem}
\emph{\textbf{Proof:}} The expressions for the cost and constraints and convexity (linearity) of them follow from Lemma 1, Lemma 2, Lemma 3, Lemma 4, and \eqref{SigmaV_pos_def2}.  \hfill $\blacksquare$
%%%%%%%%%%%%%%%%%%%%%%%%%%%%%%%%%%%%%%%%%%%%%%%
\section{Illustrative case study}
We illustrate the performance of our tools through a case study of a well-stirred chemical reactor with a heat exchanger. The reactor state, output, and controller are:
%\begingroup\makeatletter\def\f@size{9.0}\check@mathfonts
%\def\maketag@@@#1{\hbox{\m@th\normalsize\normalfont#1}}%
\begin{align*}
\left\{
\begin{array}{ll}
x_k =  \begin{pmatrix} C_0&T_0&T_w&T_m \end{pmatrix}^\top,\,\,
y_k =  x_k,\,\,
u_k=K y_k.
\end{array}
\right.
\end{align*}%\endgroup
where %$C_0$, $T_0$, $T_w$, and $T_m$ denote the concentration of the chemical product, the temperature of the product, the temperature of the jacket water of heat exchanger, and the coolant temperature, respectively.

\begin{align*}
\left\{
\begin{array}{ll}
C_0&: \text{Concentration of the chemical product},\\
T_0&: \text{Temperature of the product},\\
T_w&: \text{Temperature of the jacket water of heat exchanger},\\
T_m&: \text{Coolant temperature}.
%%C_u&: \text{Inlet concentration of reactant},\\
%%T_u&: \text{Inlet temperature},\\
%%T_{w,u}&: \text{Coolant water inlet temperature}.
\end{array}
\right.
\end{align*}
\begin{table*}[!ht]
\noindent\rule{\hsize}{1pt}
\begin{equation} \label{eq:experiment}\left\{\begin{array}{l}
A=\left(\begin{array}{cccc}
0.8353 & \hspace{-1mm}0 & \hspace{-1mm}0 & \hspace{-1mm}0 \\
0 & \hspace{-1mm}0.8324 & \hspace{-1mm}0 & \hspace{-1mm}0.0031 \\
0 & \hspace{-1mm}0.0001 & \hspace{-1mm}0.1633 & \hspace{-1mm}0 \\
0 & \hspace{-1mm}0.0280 & \hspace{-1mm}0.0172 & \hspace{-1mm}0.9320
\end{array}\right), \, B=\left(\begin{array}{ccc}
B = 0.0458 &\hspace{-1mm}0 &\hspace{-1mm}0\\
     0 &\hspace{-1mm}0.0457& \hspace{-1mm}0\\
     0 &\hspace{-1mm}0& \hspace{-1mm}0.0231\\
     0 &\hspace{-1mm}0.0007& \hspace{-1mm}0.0006
\end{array}\right),\, \Sigma^h = 0.01I_{n_y},\, x_1 \sim \mathcal{N}[\mathbf{0},10I]\\
\Sigma^w=0.1I_{n_y},\,\,L =\left(\begin{array}{cccc}
0.4884&0&0&0\\
0&0.594&0&0.0034\\
0&0.00007&0.1226&0.0001\\
0&0.0209&0.013&0.769
\end{array}\right),\,\, K =\left(\begin{array}{cccc}
-0.1237& 0& 0& 0\\
0&-0.1286&-0.0009&-0.0435\\
0&	-0.001&	-0.004&-0.0073
\end{array}\right).
\end{array}\right.
\end{equation}
\noindent\rule{\hsize}{1pt}
\vspace{-10mm}
\end{table*}
We use the discrete-time dynamics of the reactor introduced in \cite{murguia2021privacy} for the illustrative simulation study with matrices as given in \eqref{eq:experiment}. % including the designed LQR controller gain $K$ and the observer gain $L$. 
We implement the algorithm for two privacy mechanisms: first when the privacy mechanism is as in \eqref{eq2} and the second when the privacy mechanism does not include matrix transformation ($G=I$), to evaluate the effect of $G$ in privacy mechanisms.
%%%%%%%%%%%%%%%%%%%%%%%%%%%%%%%%%%%%cost
%\begin{figure}[ht]
%\begin{subfigure}{.5\textwidth}
  % include first image %2.95
 % \includegraphics[width=3.5in]{Figures/cost3.jpg}
%\end{subfigure}
%\begin{subfigure}{.5\textwidth}
  % include second image
 % \includegraphics[width=3.5in]{Figures/costwoG2.jpg}
%\end{subfigure}
%\caption{Evolution of the optimal cost function (information leakage) based on increasing $\epsilon$ for with and without matrix transformation in the privacy mechanism.}
%\end{figure}
%%%%%

\begin{figure}[!htb]
  \includegraphics[width=3.5in]{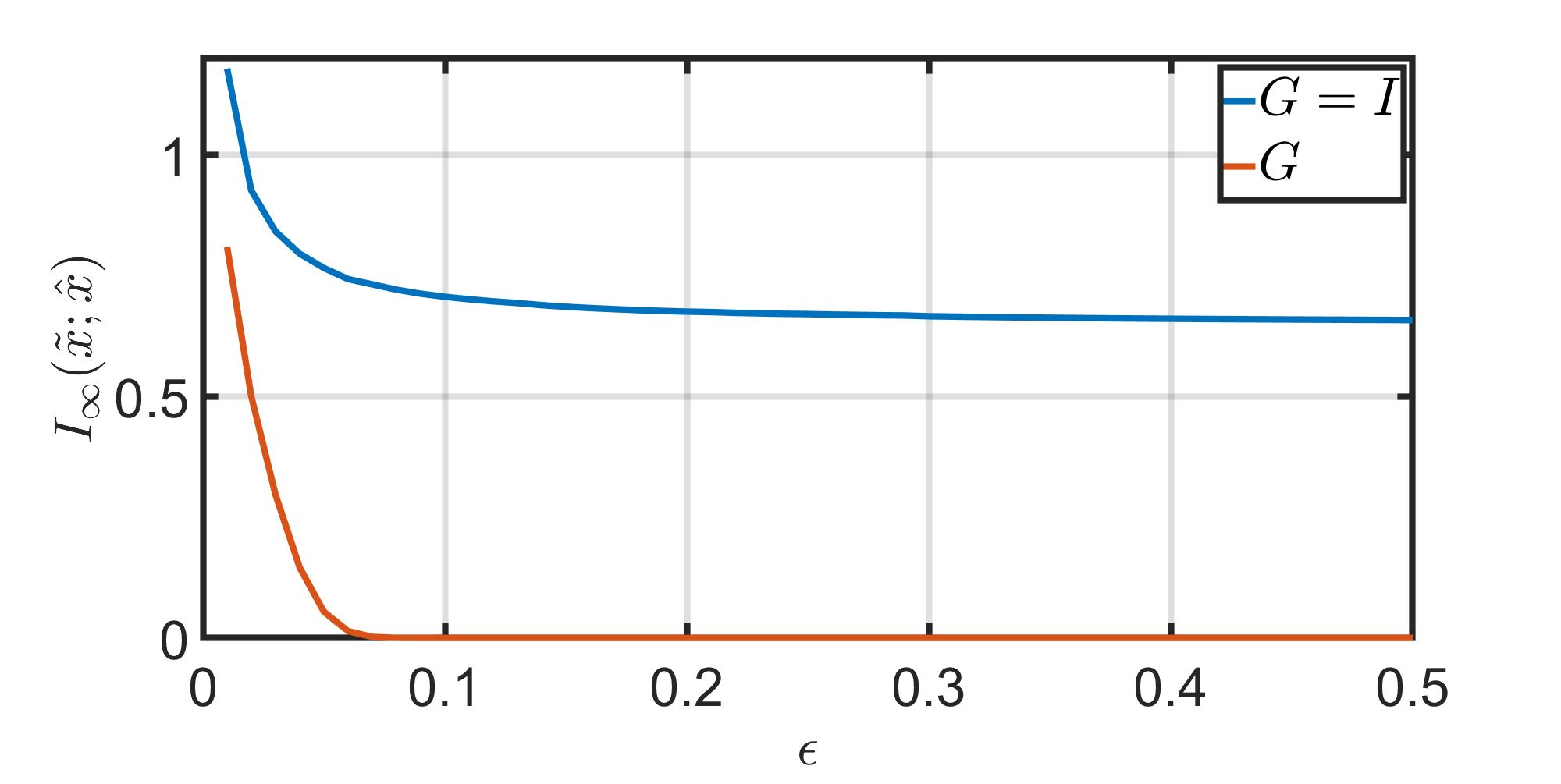}
  \caption{Evolution of the optimal cost function (information leakage) based on increasing $\epsilon$ for with and without matrix transformation in the privacy mechanism.}\label{cost}
\end{figure}

First, we show the effect of the control performance degradation level $\epsilon$ on the (mutual information-based) privacy cost function. Fig. \ref{cost} depicts the evolution of the optimal cost $I_{\infty}(\tilde{x};\hat{x})$ for increasing $\epsilon$ for both with and without matrix transformation in privacy mechanism cases shown by $G$ and $G=I$, respectively. % for two cases, the first one when the privacy mechanism is as \eqref{eq2} and the second one when the privacy mechanism does not include matrix transformation ($G=I$).
As expected, in both cases, the objective function decreases monotonically for the increased maximum allowed control performance degradation. Furthermore, given that the control cost without privacy distortion is $C_{\infty}(x,u)=4.3615$, this figure illustrates that in the case of with matrix transformation $G$, the infinite horizon optimal information leakage, which is shown by optimal $I_{\infty}(\tilde{x};\hat{x})$, can get very close to zero by a very small control performance degradation level ($\epsilon=0.07$). So, in this case, we can minimize the information leakage without degrading the control performance excessively. Hence, the comparison between the information leakage in these two cases indicates that adding matrix transformation in the privacy mechanism \eqref{eq2} improves privacy by decreasing the information leakage significantly.
%%%%%%%%%%%%%%%%%%%%%%%%%%%%%%%%%%%%%%%%%%%%%
%%%%%%%%%%%%%%%%%%%%%%%%%%%%%%%%%%%%%%control cost
%\begin{figure}[!htb]
%  \includegraphics[width=3.5in]{Figures/controlcostall.jpg}
%  \caption{Comparison between the control cost without privacy $C_{\infty}$, upper bound of the control cost distortion $C_{\infty}+\epsilon$, and the distorted control cost $\tilde{C}_{\infty}$ with and without matrix transformation for increasing $\epsilon$.}\label{controlcost}
%\end{figure}
%%%%%%%%%%%%%%%%%%%%%%%%%%%%%%%%%%%%%%%%%%%%%%%%%%%%
%
%Next, in Fig. \ref{controlcost}, we depict the LQR control cost by increasing $\epsilon$. As we expected, the control cost is increasing by increasing $\epsilon$ in both the cases with and without $G$, and in the $G=I$ case, it is fixed after a certain $\epsilon$, because, as we discussed, by increasing $\epsilon$ the Lyapunov equation \eqref{lyapanov} won't be solvable if $\mathcal{B}$, and as a result the control cost, increases.
%%%%%%%%%%%%%%%%%% 
%Then, in Fig. \ref{compare}, the evolution of optimal cost is shown in terms of increasing the LQR control cost. This figure shows the effectiveness of the proposed privacy mechanism \eqref{eq2} in minimizing information leakage even without increasing the control cost excessively.
%%%%%%%%%%%5
%%%%%%%%%%%%%%%%%%%%%%%%%%%%%%%%%%%%%% state estimate
\begin{figure}[ht]
%\begin{center}
\begin{subfigure}{.4\textwidth}
% include first image %2.95
\includegraphics[width=3.2in]{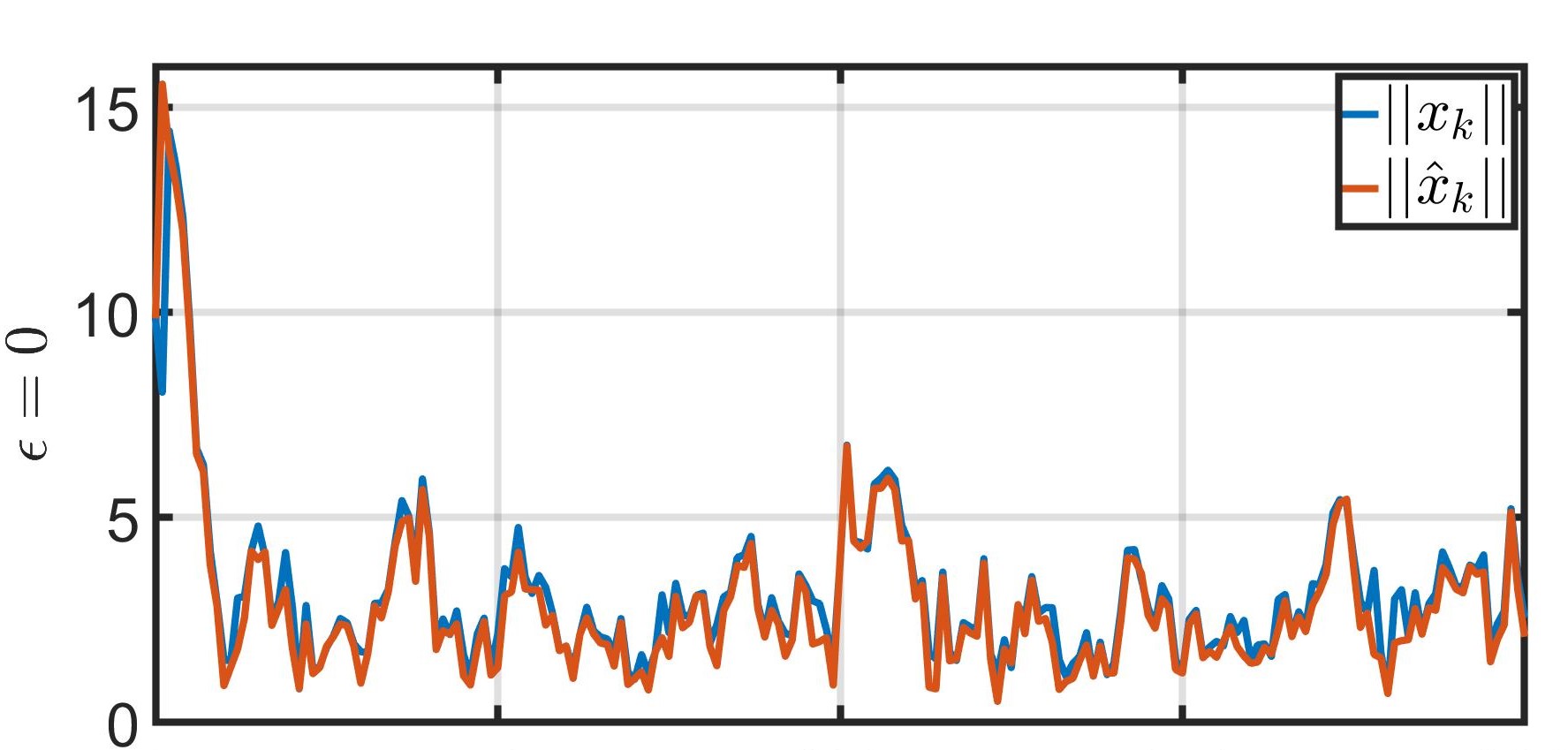}
\end{subfigure}
\begin{subfigure}{.4\textwidth}
% include second image
\includegraphics[width=3.2in]{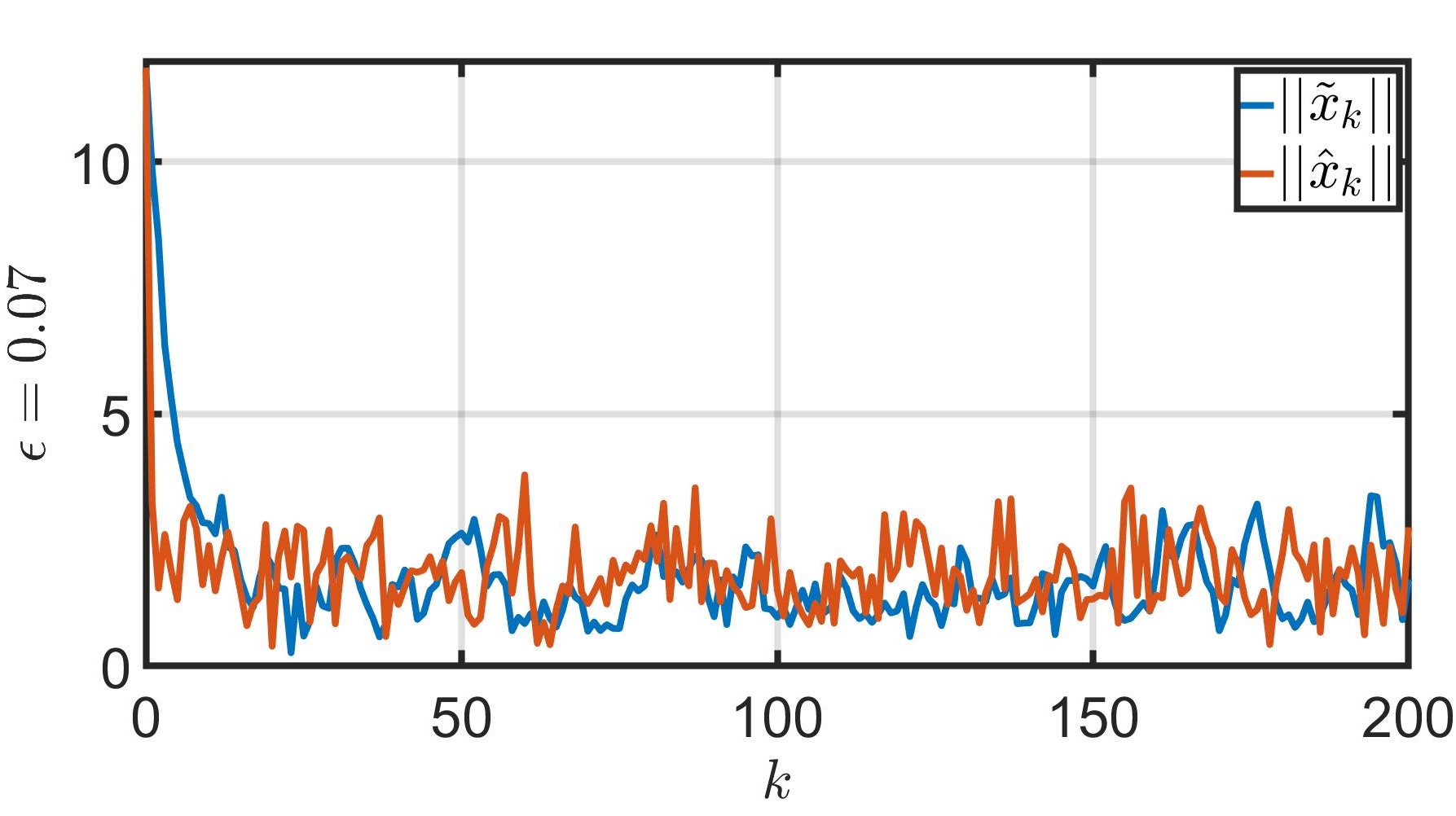}
\end{subfigure}
\caption{Comparison between the norm of system state and its Kalman estimate for $\epsilon=0,0.07$.}
\label{xestimate}
%\end{center}
\end{figure}
%\section*{Acknowledgment}
%%%%%%%%%%%%%%%%%%%%%%%%%%%%%%%%%%%%%%%%%%%%%%%%%%%%%%%%

Then, in Fig. \ref{xestimate}, we depict the norm of the system state and its Kalman estimate with and without privacy distortion. As can be seen in this figure, the accuracy of state estimation based on distorted data $(\Tilde{y}_k,\Tilde{u}_k)$ with $\epsilon=0.07$ is less than the estimation accuracy without privacy distortion ($\epsilon=0$). The mean squared error for state estimation is $4.1304$ and $1.5337$ with and without the proposed privacy solution. Therefore, we can prevent accurate estimation of the private state using the proposed privacy tools. 

\begin{figure}[!htb]
%\begin{subfigure}{.4\textwidth}
% include first image %2.95
%\includegraphics[width=3.5in]{Figures/y1001e.jpg}
%\end{subfigure}
%\begin{subfigure}{.4\textwidth}
% include second image
%\includegraphics[width=3.5in]{Figures/y102e.jpg}
%\end{subfigure}
\includegraphics[width=3.5in]{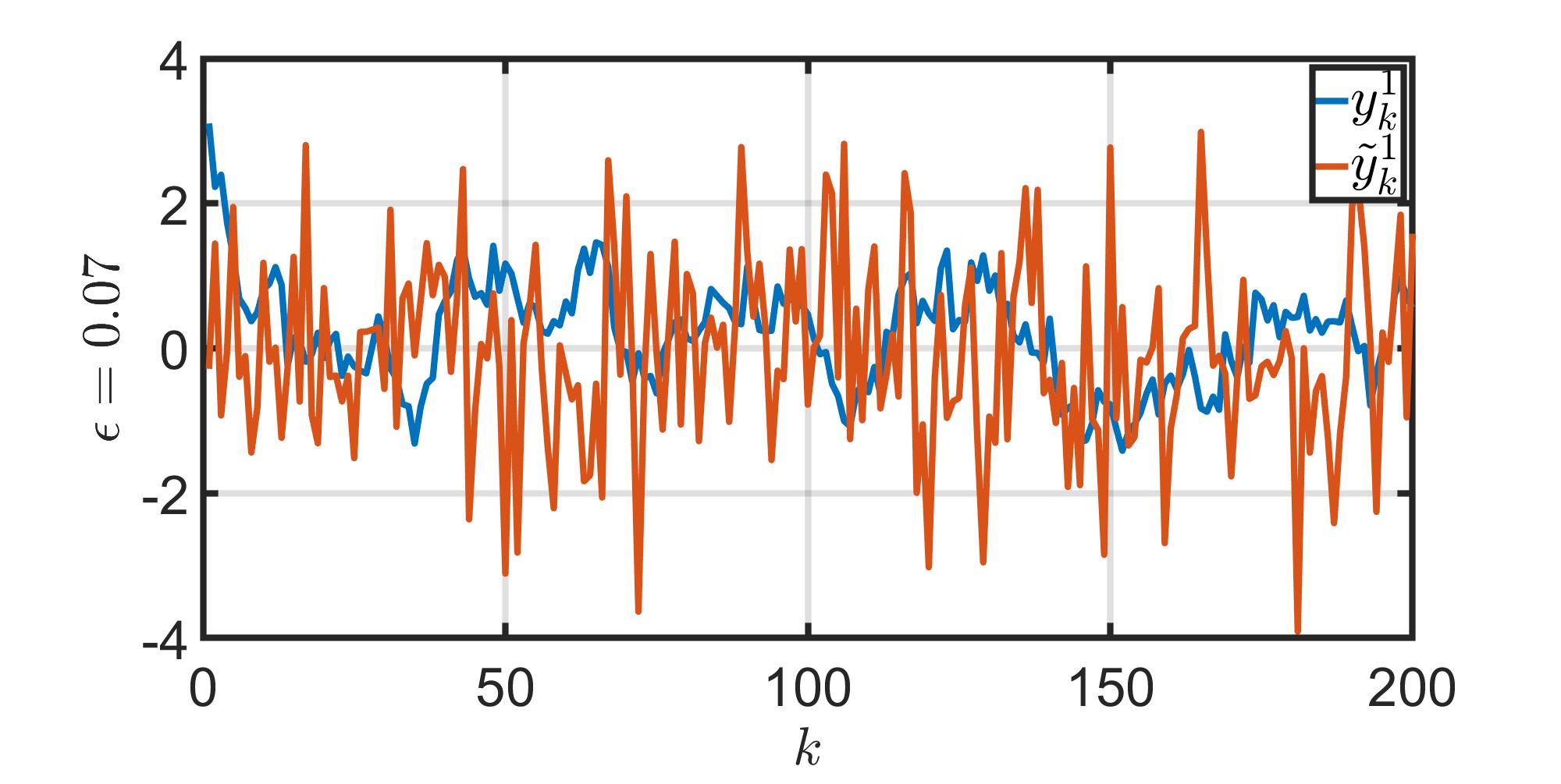}
\caption{Comparison between the first element of measurable output $y^1_k$ and the first element of the distorted output $\tilde{y}^1_k$.}\label{yyt}
\end{figure}

Finally, the effect of the optimal distortion mechanisms is illustrated in Figure \ref{yyt}, where we contrast actual and distorted measurable output for $\epsilon=0.07$.
%%%%%%%%%%%%%%%%%%%%%%%%%%%%%%%%%%%%%%control cost
%\begin{figure}[!htb]
%  \includegraphics[width=3.5in]{Figures/compare.jpg}
%  \caption{Evolution of the optimal cost function (information leakage) based on the increasing control cost for with and without matrix transformation in the privacy mechanism.}\label{compare}
%\end{figure}
%%%%%%%%%%%%%%%%%%%%%%%%%%%%%%%%%y and ytilda
%%%%%%%%%%%%%%%%%%%%%%%%%%%%%%%%%%%%%%%%
\section{Conclusions}
In this paper, for a class of Networked Control Systems (NCSs), we have presented a detailed mathematical framework for synthesizing distorting mechanisms to minimize the infinite horizon information leakage induced by the use of public/unsecured communication networks. We have proposed a class of linear Gaussian distorting mechanisms to randomize sensor and control data before transmission to prevent adversaries from accurately estimating the system state. Furthermore, for the class of systems under study, we have fully characterized an information-theoretic metric (mutual information) to quantify the information between the system state and its optimal estimate given the distorted disclosed data at the remote station for a class of worst-case eavesdropping adversaries. Finally, given the maximum allowed level of control performance degradation (LQR cost), we have provided tools (in terms of convex programs) to design sub-optimal (in terms of maximizing privacy) distorting mechanisms. We have presented simulation results to illustrate the performance of our tools.
%%%%%%%%%%%%%%%%%%%%%%%%%%%%%%%%%%%%%%%%%%%%%%%%%%%%%%%%%%%%%
\section{Acknowledgment}
The research leading to these results has received funding from the European Union’s Horizon Europe programme under grant agreement No 101069748 – SELFY project.
%%%%%%%%%%%%%%%%%
\section{Appendix}
%%%%%%%%%%%%%%%%%%%%%%%%%%%%%%%%%%%%%%%%%%%%%%%
%%%%%%%%%%%%%%%%%%%%%%%%%%%%%%%%%%%%%%%%%%%%%%%
\subsection{Proof of Lemma 1}
\noindent
The uplink information flow $I\left(\tilde{x}^N \rightarrow \hat{x}^N\right)$ is given by \cite{massey1990causality}:
\begin{equation}
\begin{aligned}
I\left(\tilde{x}^N \rightarrow \hat{x}^N\right)&=\sum_{k=0}^N I\left(\tilde{x}^k ; \hat{x}_k \mid \hat{x}^{k-1}\right). %\\
%&=\sum_{k=0}^NI\left(x^{k-1}, x_k ; \hat{x}_k \mid \hat{x}^{k-1}\right).
\end{aligned}
\end{equation}
Then, based on the chain rule in mutual information \cite{Cover}:
\begin{equation}\label{xxx}
\begin{aligned}
 &I\left(\tilde{x}^N \rightarrow \hat{x}^N\right)\\
 &\,\,\, =\sum_{k=1}^N[\underbrace{I\left(\tilde{x}^{k-1} ; \hat{x}_k \mid \hat{x}^{k-1}, \tilde{x}_k\right)}_{=(A)}+\underbrace{I\left(\tilde{x}_k ; \hat{x}_k \mid \hat{x}^{k-1}\right)}_{=(B)}].
 \end{aligned}
\end{equation}
By substituting $\tilde{y}_k$ in \eqref{eq1priv} into \eqref{kalmanfilter}, we have $\hat{x}_k$ in terms of $\hat{x}_{k-1}$, $\tilde{x}_k$, and noises as follows:
\begin{equation}\label{xhat}
    \hat{x}_k=(I-L) A\hat{x}_{k-1}+(I-L) Bu_{k-1}+L G \tilde{x}_k+L\Tilde{v}_k.
\end{equation}
%where $\Tilde{v}_k:=Gh_k+v_k$ and $\Sigma^{\Tilde{v}}:=G\Sigma^hG^\top + \Sigma^V$.
Then, considering \eqref{xhat} and the fact that $u_{k-1}=K y_{k-1}$ is a deterministic function of $\hat{x}^{k-1}$ (see \eqref{kalmanfilter}), we have:
\begin{equation}\label{Ais0}
(A)=I\left(\tilde{x}^{k-1} ; L\Tilde{v}_k \mid \hat{x}_{k-1}, x_k\right)=0.
\end{equation}
Substituting \eqref{xhat} in (B) and using mutual information definition in terms of differential entropy \cite{Cover}, we have
\begin{equation}\label{000}\resizebox{.49 \textwidth}{!}{
$\begin{aligned}
(B)&=I\left(\tilde{x}_k ; (I-L) Bu_{k-1}+L G \tilde{x}_k+L\Tilde{v}_k \mid \hat{x}^{k-1}\right) \\
& =\underbrace{h\left(L G \tilde{x}_k+L\Tilde{v}_k  \mid \hat{x}^{k-1}\right)}_C-\underbrace{h\left(L G \tilde{x}_k+L\Tilde{v}_k  \mid \hat{x}^{k-1}, \tilde{x}_k\right)}_D.
\end{aligned}$}
\end{equation}
%%%%%%%%
Given the system dynamics \eqref{eq1priv} and the fact that the estimation error $e_{k-1}$ is independent of the previous measurement (and of $\hat{x}^{k-1}$ ), and by substituting $e_{k|k-1}$ given in \eqref{estimationerrordynamics}, (C) is simplified as follows:
\begin{equation}\label{1112}\resizebox{.49 \textwidth}{!}{
$\begin{aligned}
(C) & =h\left(L G\left(A \tilde{x}_{k-1}+ B\tilde{u}_{k-1} +w_{k-1}\right)+L\Tilde{v}_k \mid \hat{x}^{k-1}\right) \\
& =h\left(LG\left( A e_{k-1}+ Bz_{k-1}+ w_{k-1}\right)+L\Tilde{v}_k\mid \hat{x}^{k-1}\right) \\
&=\frac{1}{2} \log \operatorname{det}\left(LG \Sigma_{k|k-1}^e G^\top L^\top +L \Sigma^{\Tilde{v}} L^{\top}\right).
\end{aligned}$}
\end{equation}
%From \eqref{estimationerrordynamics}, %$\hat{x}_{k|k-1}=A\hat{x}_{k-1}+Bu_{k-1}$ and as a result, $\Sigma_{k|k-1}^e = A \Sigma_{k-1}^e A^{\top}+B \Sigma^{z} B^{\top}  +\Sigma^{w}$, 
%it follows that:
%\begin{equation}\label{111}
% (C) =   \frac{1}{2} \log \operatorname{det}\left(LG \Sigma_{k|k-1}^e G^\top L^\top +L \Sigma^{\Tilde{v}} L^{\top}\right).
%\end{equation}
Also, because $\tilde{v}_k$ is i.i.d., (D) can be written as follows:
\begin{eqnarray}\label{222}
\begin{aligned}
(D)=h\left(L\Tilde{v}_k\mid\hat{x}^{k-1}, \tilde{x}_k\right)=\frac{1}{2} \log \operatorname{det}\left(L \Sigma^{\Tilde{v}} L^{\top}\right).
\end{aligned}
\end{eqnarray}
Therefore, substituting \eqref{Ais0}, \eqref{000}, \eqref{1112}, and \eqref{222} into \eqref{xxx}, the uplink directed information is calculated as follows:
\begin{equation}\label{uplinkinf}
\begin{aligned}
I\left(\tilde{x}^N \rightarrow \hat{x}^N\right)&=\sum_{k=1}^N\left(\frac{1}{2} \log \operatorname{det}\left(LG \Sigma_{k|k-1}^e G^\top L^\top\right.\right. \\
&\left.\left.+L \Sigma^{\Tilde{v}} L^{\top}\right)  -  \frac{1}{2} \log \operatorname{det}\left(L\Sigma^{\Tilde{v}} L^{\top}\right)\right).
\end{aligned}
\end{equation}
Following the same procedure, the downlink directed information can be written as follows:
\begin{equation}\label{downlinkinf}\resizebox{.5 \textwidth}{!}{
$\begin{aligned}
&I\left(0 * \hat{x}^{K-1} \rightarrow \tilde{x}^{K}\right)=I\left(0 ; \tilde{x}_1 \mid \tilde{x}_0\right)+\sum_{k=1}^N I\left(\hat{x}^{k-1} ; \tilde{x}_k \mid \tilde{x}^{k-1}\right) \\
&\,\,\, =\sum_{k=1}^N\left(h\left(\tilde{x}_k \mid \tilde{x}^{k-1}\right)-h\left(\tilde{x}_k \mid \hat{x}^{k-1}, \tilde{x}^{k-1}\right)\right) \\
&\,\,\, =\sum_{k=1}^N\left(h\left(BK \Tilde{v}_{k-1}+Bz_{k-1} +w_{k-1}\right)-h\left(Bz_{k-1}+w_{k-1}\right)\right)\\
&\,\,\, =\sum_{k=1}^N\left(\frac{1}{2} \log \operatorname{det}\left(BK \Sigma^{\Tilde{v}}K^{\top}B^\top +B \Sigma^zB^\top + \Sigma^w\right)\right) \\&\qquad- \left( \frac{1}{2} \log \operatorname{det}\left(B \Sigma^zB^\top + \Sigma^w\right)\right).\\
\end{aligned}$}
\end{equation}
Therefore, $ I\left(\tilde{x}^N ; \hat{x}^N\right)$ is calculated by the summation of uplink \eqref{uplinkinf} and downlink \eqref{downlinkinf} information flows.
\hfill $\blacksquare$
%%%%%%%%%%%%%%%%%%%%%%%%%%%%%%%%%%%%%%%%%
%%%%%%%%%%%%%%%%%%%%%%%%%%%%%%%%%%%%%%%%%%%%%%%%%%%%%%%%%%%%%%%%%%%%%%%%%%%%%%%%%%%%%%%%%%%%%%%%%%%%%%%%
%%%%%%%%%%%%%%%%%%%%%%%%%%%%%%%%%%%%%%%%%%%%%%%
\subsection{Proof of Lemma 2}
At first, we prove that an upperbound for the solution of $E_1:=\mathcal{A} \Sigma^{\zeta} \mathcal{A}^\top -\Sigma^{\zeta} +\mathcal{B}=\mathbf{0}$, can be achieved by solving:
\begin{equation}\label{extra00}
\left\{\begin{aligned}
&\min_{\Sigma} \operatorname{trace}(\Sigma),\\[1mm]
    &\hspace{4mm}\text{s.t. }
E_2:=\mathcal{A}\Sigma \mathcal{A}^\top- \Sigma + \mathcal{B} \le \mathbf{0}.
\end{aligned}\right.
\end{equation}
%Given that $\mathcal{A}$ is Schur stable, it can be proved that an upperbound of the solution of the DARE \eqref{Riccatiricursion} can be calculated by the inequality $\mathcal{A}\Sigma \mathcal{A}^\top- \Sigma + \mathcal{B} \le \mathbf{0}$. 
From $E_2 \le E_1$, it can be deduced that:
\begin{equation}\label{extra0}
    \mathcal{A}(\Sigma-\Sigma^{\zeta} )\mathcal{A}^\top- (\Sigma-\Sigma^{\zeta} )\le \mathbf{0}.
\end{equation}
Then, from \eqref{extra0} and given that $\mathcal{A}$ is Schur stable, we can conclude that $\Sigma\ge \Sigma^{\zeta}$. Hence, minimizing $\operatorname{trace}(\Sigma)$ with inequality $E_2 \le \mathbf{0}$ as the constraint gives us an upperbound on $\Sigma^{\zeta}$ which is the solution of Lyapanov equation \eqref{Riccatiricursion}.\\
Using standard Schur complement properties \cite{zhang2006schur}, the nonlinear inequality $E_2\le \mathbf{0}$ can be converted to:
\begin{equation}\label{pr21}
\left[\begin{array}{cc}
\Sigma-\mathcal{B} & \mathcal{A} \\
{\mathcal{A}}^{\top} & \Sigma^{-1}
\end{array}\right] \ge \mathbf{0}.
\end{equation}
We define an invertible matrix $\Pi_1$ as a new design variable as in \eqref{pr0}. %follows:
%\begin{equation}\label{Pi1def}
%    \Pi_1:=\left[\begin{array}{cc}
% \Pi_{11}&\Pi_{12}\\
% \mathbf{0}&\Pi_{13}
%\end{array}\right].
%\end{equation}
It follows that a congruence transformation of \eqref{pr21} can be written as follows, that is positive definite since \eqref{pr21} is positive definite \cite{BEFB:94}:
\begin{equation}\label{pr22}
\left[\begin{array}{cc}
I & \mathbf{0} \\
\mathbf{0} & \Pi_1
\end{array}\right]^{\top} \left[\begin{array}{cc}
\Sigma-\mathcal{B} & \mathcal{A} \\
{\mathcal{A}}^{\top} & \Sigma^{-1}
\end{array}\right] \left[\begin{array}{cc}
I & \mathbf{0} \\
\mathbf{0} & \Pi_1
\end{array}\right] \ge \mathbf{0},
\end{equation}
which is equivalent to
\begin{equation}\label{pr23}
\left[\begin{array}{cc}
\Sigma-\mathcal{B} & \mathcal{A} \Pi_1 \\
(\mathcal{A}\Pi_1)^{\top} & \Pi_1^{\top} \Sigma^{-1}\Pi_1 
\end{array}\right] \ge \mathbf{0}.
\end{equation}
It can be proved that, for any matrix $\bar{A}$ and invertible matrix $\bar{B}$ we have (Hint: $\left(\bar{B}^{-1/2}\bar{A} -\bar{B}^{1/2}\right)^\top \left(\bar{B}^{-1/2}\bar{A} -\bar{B}^{1/2}\right) \ge \mathbf{0}$):
\begin{equation}\label{invrule}
    \bar{A}^\top \bar{B}^{-1} \bar{A} \ge \bar{A}+\bar{A}^\top-\bar{B}.
\end{equation}
Therefore, from \eqref{pr23} and \eqref{invrule}, we can conclude that:
\begin{equation}\label{pr24}
\left[\begin{array}{cc}
\Sigma-\mathcal{B} & \mathcal{A} \Pi_1 \\
(\mathcal{A}\Pi_1)^{\top} & \Pi_1^{\top} + \Pi_1 -\Sigma 
\end{array}\right] \ge \mathbf{0}.
\end{equation}
Matrix $\mathcal{A}$, given in \eqref{mathcalAB}, can be written as follows:
\begin{equation}\label{AA}
    \mathcal{A}=\mathcal{A}_0 + \mathcal{A}_1 G \left[\begin{array}{cc}
\mathbf{0} &I
\end{array}\right],
\end{equation}
where $\mathcal{A}_{0}$ and $\mathcal{A}_1$ are defined in \eqref{definematrices}.
%\begin{equation}
%    \mathcal{A}_{0}:=\left[\begin{array}{cc}
%A(I-L) & AL \\
%\mathbf{0} & A
%\end{array}\right],\hspace{4mm} \mathcal{A}_1:=\left[\begin{array}{c}
%-AL \\BK
%\end{array}\right].
%\end{equation}
By defining new design variable $\Pi_2:= G \left[\begin{array}{cc}
\mathbf{0} &I
\end{array}\right] \Pi_1=\left[\begin{array}{cc}
\mathbf{0} &G \Pi_{13}
\end{array}\right]=\left[\begin{array}{cc}
\mathbf{0} &\Pi_{21}
\end{array}\right]$ and substituting \eqref{AA} into \eqref{pr24}, \eqref{extra00} can be converted to \eqref{pr0}, which is linear in design variables $\Pi_1$, $\Pi_2$, and $\Sigma$.
\hfill $\blacksquare$
%%%%%%%%%%%%%%%%%%%%%%%%%%%%%%%%%%%%%%%%%%%%%%%%%%%%%%%%%%%%%%%%%%%%%%%
\subsection{Proof of Lemma 3}
Due to the monotonicity of the logarithm determinant function, minimizing the right-hand side of \eqref{costf} is equivalent to solving the following optimization problem:
\begin{equation}
\left\{\begin{aligned}
	&\min _{\Sigma^{\Tilde{v}},G,\Sigma^z,\Sigma,\Pi_3,\Pi_4} 
 \frac{1}{2} \log \operatorname{det} \left(\Pi^{-1}_3\right)- \frac{1}{2} \log \operatorname{det}\left(L\Sigma^{\Tilde{v}} L^{\top}\right)\\&\qquad+\frac{1}{2} \log \operatorname{det}\left(\Pi^{-1}_4\right) - \frac{1}{2} \log \operatorname{det}\left(B \Sigma^zB^\top + \Sigma^w\right)\label{convexcost1}\\[1mm]
    &\text{\emph{s.t. }} 
    \left\{\begin{aligned}
    & \Pi^{-1}_3 \ge \left(L\left(G \bar{\Sigma}^e G^\top +\Sigma^{\Tilde{v}}\right) L^{\top}\right),\\
    &\Pi^{-1}_4 \ge \left(BK \Sigma^{\Tilde{v}}K^\top B^\top +B \Sigma^zB^\top + \Sigma^w \right),
    \end{aligned}\right.
\end{aligned}\right.
\end{equation}
where $\bar{\Sigma}^e:=N_{e} \Sigma N_{e}^\top \ge {\Sigma}^e$. From relation \eqref{invrule}, we can conclude $\Pi^{-1}_3 \ge 2I - \Pi_3$ and $\Pi^{-1}_4 \ge 2I - \Pi_4$ which linearizes the second inequality term of \eqref{convexcost1}. Then, the first inequality term of \eqref{convexcost1} is equivalent to its Schur complement as follows \cite{zhang2006schur}:
\begin{equation}\label{pr301}
\left[ {\begin{array}{*{20}{c}}
2I - \Pi_3 - L\Sigma^{\Tilde{v}}L^\top & LG\\
(LG)^\top& (\bar{\Sigma}^e)^{-1}
\end{array}} \right] \ge \mathbf{0}.
\end{equation}
A congruence transformation of \eqref{pr301} can be calculated as follows \cite{BEFB:94}:
\begin{equation}\label{pr31}\resizebox{.5 \textwidth}{!}{
$\left[\begin{array}{cc}
I & \mathbf{0} \\
\mathbf{0} & \Pi_{13}
\end{array}\right]^{\top} \left[ {\begin{array}{*{20}{c}}
2I - \Pi_3 - L\Sigma^{\Tilde{v}}L^\top & LG\\
(LG)^\top& (\bar{\Sigma}^e)^{-1}
\end{array}} \right] \left[\begin{array}{cc}
I & \mathbf{0} \\
\mathbf{0} & \Pi_{13}
\end{array}\right] \ge \mathbf{0}.$}
\end{equation}
By relation \eqref{invrule}, we have $\Pi_{13}^{\top} (\bar{\Sigma}^e)^{-1} \Pi_{13} \ge \Pi_{13}^{\top} +\Pi_{13} - \bar{\Sigma}^e$. Then, given $G\Pi_{13}=\Pi_{21}$, \eqref{pr31} can be converted to:
\begin{equation}\label{pr32}\left[ {\begin{array}{*{20}{c}}
2I - \Pi_3 - L\Sigma^{\Tilde{v}}L^\top &  L \Pi_{21}\\
*& \Pi_{13}+\Pi_{13}^\top -\bar{\Sigma}^e
\end{array}} \right] \ge \mathbf{0}.
\end{equation}
Combining \eqref{convexcost1} and \eqref{pr32}, an upperbound for the optimal value of $I_{\infty}(\tilde{x};\hat{x})$ in \eqref{costf} can be achieved by solving the convex program in \eqref{inequalityofcost1}. 
\hfill $\blacksquare$
%%%%%%%%%%%%%%%%%%%%%%%%%%%%%%%%%%%%%%%%%%%%%%%%%%%%%%%%%%
%%%%%%%%%%%%%%%%%%%%%%%%%%%%%%%%%%%%%%%%%%%%%%%%%%%%%%%%%%%%%%%%%%%%%%%
\subsection{Proof of \eqref{SigmaV_pos_def2}}
A congruence transformation of \eqref{SigmaV_pos_def} can be written as follows \cite{BEFB:94}:%, which is positive definite since \eqref{SigmaV_pos_def} is positive definite \cite{BEFB:94}:
\begin{equation}\label{pr001}
\left[\begin{array}{cc}
I & \mathbf{0} \\
\mathbf{0} & \Pi_{13}
\end{array}\right]^{\top} \left[ {\begin{array}{*{20}{c}}
\Sigma^{\Tilde{v}}    &   G \\
{G}^\top      &    (\Sigma^h)^{-1}
\end{array}} \right] \left[\begin{array}{cc}
I & \mathbf{0} \\
\mathbf{0} & \Pi_{13}
\end{array}\right] \ge \mathbf{0},
\end{equation}
which is equivalent to
\begin{equation}\label{pr002}
\left[ {\begin{array}{*{20}{c}}
\Sigma^{\Tilde{v}}    &   G\Pi_{13} \\
(G\Pi_{13})^\top      &    \Pi_{13}^{\top}(\Sigma^h)^{-1} \Pi_{13}
\end{array}} \right] \ge \mathbf{0}.
\end{equation}
By relation \eqref{invrule}, we have $\Pi_{13}^{\top}(\Sigma^h)^{-1} \Pi_{13} \ge \Pi^{\top}_{13}+\Pi_{13} -\Sigma^h$. Then, given that $G\Pi_{13}=\Pi_{21}$, equation \eqref{pr002} is converted to \eqref{SigmaV_pos_def2}.
\hfill $\blacksquare$
%%%%%%%%%%%%%%%%%%%%%%%%%%%%%%%%%%%%%%%%%%%%%%%%%%%%%%%%%%
\subsection{Proof of Lemma 4}
We define $\Delta:= \tilde{x}_k^\top Q \tilde{x}_k +\tilde{u}_k^\top R \tilde{u}_k$. Then, given the distortion mechanism \eqref{eq2} and system dynamics \eqref{eq1priv}, $\Delta$ can be calculated as follows:
\begin{equation}\label{pr42}
\begin{aligned}
\Delta & =\tilde{x}_k^{\top}\left(Q+G^{\top} K^{\top} R K G\right) \tilde{x}_k+\Tilde{v}_k^{\top} K^{\top} R K \Tilde{v}_k\\
&+\Tilde{v}_k^{\top}K^{\top} R K G x_k  +\tilde{x}_k^{\top} G^{\top} K^{\top} R K \Tilde{v}_k\\
&+z_k^\top Rz_k+ \tilde{x}_k^\top G^\top K^\top R z_k+\tilde{v}_k^\top K^\top R z_k\\
&+z_k^\top RKG \tilde{x}_k +z_k^\top RK \tilde{v}_k.
\end{aligned}
\end{equation}
The expectation of the quadratic form of any variable $p$ in terms of its mean $\mu^p$ and covariance $\Sigma^p$ can be calculated as $ \mathbb{E}\left[p^\top A p\right]=  \text{tr}[A\Sigma^p] + (A\mu^{p})^\top (A\mu^p) $ (see \cite{seber2012linear} for details). Then, in \eqref{pr42}, since $\tilde{x}_k$, $\tilde{v}_k$, and $z_k$ are independent, and they have zero mean, $\mathbb{E}\left[\Delta\right]$ can be calculated as follows:
\begin{equation}
\begin{aligned}
\mathbb{E}\left[\Delta\right]&= \text {tr}\left(\left(Q+G^{\top}{K^{\top}} R K G\right) \Sigma^{\tilde{x}}_k+\left(K^{\top} R K \right)\Sigma^{\Tilde{v}}\right.\\
&+\left.R\Sigma^z\right).
\end{aligned}
\end{equation}\label{pr43}
Given the upperbound $\bar{\Sigma}^{\tilde{x}}:= N_{\tilde{x}} \Sigma N_{\tilde{x}}^\top \ge \lim _{k \rightarrow \infty} \Sigma^{\tilde{x}}_{k}$ in \eqref{estimationerrorcov} and the fact that trace is a linear mapping, for $\tilde{C}_{\infty}(\tilde{x}, \tilde{u})$ we have:
\begin{equation}
\begin{aligned}
    &\tilde{C}_{\infty}(\tilde{x}, \tilde{u}) = \limsup _{K \rightarrow \infty} \frac{1}{N+1} \sum_{k=0}^N \mathbb{E}\left[\Delta \right]\\
    &\le \text {tr}\left(\left(Q+G^{\top}{K^{\top}} R K G\right) \bar{\Sigma}^{\tilde{x}}\right)+\operatorname{tr}\left(K^{\top} R K \Sigma^{\Tilde{v}}+R\Sigma^z\right).
\end{aligned}
\end{equation}
Up to this point, we have written an upperbound for $\tilde{C}_{\infty}(\tilde{x},\tilde{u})$ in terms of the design variables. Then, the constraint \eqref{constraint} is equivalent to:
\begin{equation}\label{pr44}
\begin{aligned}
       & \text {tr}\left(\left(Q+G^{\top}{K^{\top}} R K G\right) \bar{\Sigma}^{\tilde{x}}\right)\\&\qquad+\operatorname{tr}\left(K^{\top} R K \Sigma^{\Tilde{v}}+R\Sigma^z\right) \le C_{\infty}(x, u)+\epsilon.
 \end{aligned}
\end{equation}
Since trace is a linear mapping, \eqref{pr44} can be converted to:
\begin{equation}\label{pr45}
\begin{aligned}
       & \text {tr}\left(Q \bar{\Sigma}^{\tilde{x}}\right)+\text {tr}\left(R^{1/2}K G \bar{\Sigma}^{\tilde{x}} G^\top K^\top (R^{1/2})^\top\right)\\
       &\qquad+\operatorname{tr}\left(K^{\top} R K \Sigma^{\Tilde{v}}+R\Sigma^z\right) \le C_{\infty}(x, u)+\epsilon.
 \end{aligned}
\end{equation} 
However, due to the monotonicity of the trace function, the inequality \eqref{pr45} can be converted to the following set of inequalities:
\begin{subequations}
\begin{eqnarray}
&\text {tr}\left(Q\bar{\Sigma}^{\tilde{x}}\right)+\text {tr}\left(\Pi_5\right)\nonumber \\&\qquad+\operatorname{tr}\left(K^{\top} R K \Sigma^{\Tilde{v}}+R\Sigma^z\right)\le C_{\infty}(x, u)+\epsilon,\label{pr451}\\
&\Pi_5 \ge R^{1/2}K G \bar{\Sigma}^{\tilde{x}} G^\top K^\top (R^{1/2})^\top.\label{pr452}
\end{eqnarray}
\end{subequations}
The inequality \eqref{pr451} is linear in design variables, and \eqref{pr452} is equivalent to its Schur compliment as follows \cite{zhang2006schur}:
\begin{equation}\label{pr46}
\left[ {\begin{array}{*{20}{c}}
\Pi_5 & R^{1/2}K G\\
*& (\bar{\Sigma}^{\tilde{x}})^{-1}
\end{array}} \right] \ge \mathbf{0}.
\end{equation}
Then, a congruence transformation of \eqref{pr46} can be written as follows \cite{BEFB:94}:
\begin{equation}\label{ee}
\left[\begin{array}{cc}
I & \mathbf{0} \\
\mathbf{0} & \Pi_{13}
\end{array}\right]^{\top} \left[ {\begin{array}{*{20}{c}}
\Pi_5 & R^{1/2}K G\\
*& (\bar{\Sigma}^{\tilde{x}})^{-1}
\end{array}} \right] \left[\begin{array}{cc}
I & \mathbf{0} \\
\mathbf{0} & \Pi_{13}
\end{array}\right] \ge \mathbf{0}.
\end{equation}
By relation \eqref{invrule}, we have $\Pi_{13}^{\top} (\bar{\Sigma}^{\tilde{x}})^{-1} \Pi_{13} \ge\Pi_{13}^{\top} +\Pi_{13} - \bar{\Sigma}^{\tilde{x}}$. Then, given that $G\Pi_{13}=\Pi_{21}$, \eqref{ee} is equivalent to the following inequality:
\begin{equation}\label{pr48}
\left[ {\begin{array}{*{20}{c}}
\Pi_5 &  R^{1/2}K \Pi_{21}\\
*& \Pi_{13}+\Pi_{13}^\top - \bar{\Sigma}^{\tilde{x}}
\end{array}} \right] \ge \mathbf{0}.
\end{equation}
From \eqref{pr451} and \eqref{pr48}, we can conclude that the constraint \eqref{constraint} can be formulated by the set of LMIs in \eqref{pr30}.
\hfill $\blacksquare$
%%%%%%%%%%%%%%%%%%%%%%%%
%%%%%%%%%%%%%%%%%%%%%%%%%%%%%%%%%%%%%%%%%%%%%%%%%%%%%%%%%%%%%
\bibliographystyle{IEEEtran}

\bibliography{InfiniteHorizonPrivacyforNetworkedControlSytems}
\end{document}